\documentclass{article}
\pdfoutput=1
\usepackage{arxiv}

\usepackage[utf8]{inputenc} 
\usepackage[T1]{fontenc}    
\usepackage[numbers]{natbib}
\usepackage{hyperref}       
\usepackage{url}            
\usepackage{booktabs}       
\usepackage{amsfonts}       
\usepackage{nicefrac}       
\usepackage{microtype}      
\usepackage{lipsum}
\usepackage{fancyhdr}       
\usepackage{graphicx}       
\graphicspath{{media/}}     

\usepackage{enumitem}   
\usepackage{framed,multirow}
\usepackage[english]{babel}
\renewcommand{\appendixname}{Appendix}
\usepackage{xcolor}
\usepackage{graphicx,float}
\usepackage{subfig}
\usepackage{amsmath, amsthm, amssymb, amsfonts}
\usepackage[makeroom]{cancel} 
\usepackage[boxed,ruled]{algorithm2e} 
\usepackage{tikz}
\usepackage{cleveref}
\usepackage[toc,page]{appendix}
\usetikzlibrary{positioning}
\usetikzlibrary{arrows}
\definecolor{Red}{rgb}{1,0,0}
\definecolor{Blue}{rgb}{0,0,1}

\def\bp{\mbox{\bf p}}

\def\bj{\mbox{\bf J}}
\def\bm{\mbox{\bf m}}

\def\by{\mbox{\bf y}}
\def\bA{\mbox{\bf A}}
\def\bB{\mbox{\bf B}}
\def\bC{\mbox{\bf C}}

\def\bJ{\mbox{\bf J}}
\def\bH{\mbox{\bf H}}

\def\bP{\mbox{\bf P}}

\def\bZO{\mbox{\bf 0}}

\def\cL{\mathcal{L}}

\def\cU{\mathcal{U}}
\def\cLN{\mathcal{LN}}
\def\cN{\mathcal{N}}
\def\bD{\mathcal{D}}

\def\dR{\mathbb{R}}
\def\bphi{\boldsymbol{\phi}}

\def\bSigma{\boldsymbol{\Sigma}}
\def\balpha{\boldsymbol{\alpha}}

\def\bmu{\boldsymbol{\mu}}

\def\pr{\mbox{p}}

\DeclareMathOperator*{\argmax}{arg\,max}

\def\balphaMAP{\boldsymbol{\alpha}^{\text{map}}}
\def\alphaiMAP{\alpha_i^{\text{map}}}
\def\alphaMAP{\alpha^{\text{map}}}

\def\bDk{{\mbox{\bf D}}^{(k)}}

\def\ak{a^{(k)}}
\def\wk{w^{(k)}}
\def\vik{v_i^{(k)}}
\def\vjk{v_j^{(k)}}
\def\bmk{\bm^{(k)}}
\def\bPk{\bP^{(k)}}
\def\bphia{\bphi_{\alpha}}
\def\bphima{\bphi_{\text{-}\alpha}}

\def\vibar{\bar{v}_i}

\def\bmuk{\bmu^{(k)}}
\def\bSigmak{\bSigma^{(k)}}
\def\bmmak{\bm_{\text{-}\alpha}^{(k)}}
\def\bPmak{\bP_{\text{-}\alpha}^{(k)}}
\def\bmak{\bm_{\alpha}^{(k)}}
\def\bPak{\bP_{\alpha}^{(k)}}
\def\bBak{\bB_{\alpha}^{(k)}}

\def\btilmumak{\tilde{\bmu}_{\text{-}\alpha}^{(k)}}
\def\btilSigmamak{\tilde{\bSigma}_{\text{-}\alpha}^{(k)}}
\def\bmuak{\bmu_{\alpha}^{(k)}}
\def\bmumak{\bmu_{\text{-}\alpha}^{(k)}}
\def\bSigmaak{\bSigma_{\alpha}^{(k)}}
\def\bSigmamak{\bSigma_{\text{-}\alpha}^{(k)}}
\def\bCk{\bC^{(k)}}
\def\bCkT{(\bC^{(k)})^T}
\def\bDkT{(\bDk)^T}
\def\Piik{P_{ii}^{(k)}}

\def\gammaisum{\gamma_i^{\text{rms}}}  
\def\gammasum{\gamma^{\text{rms}}} 
\def\gammaik{\gamma_{i}^{(k)}}
\def\gammatol{\gamma^{\text{tol}}}
\def\mik{m_{i}^{(k)}}

\def\Ny{N_{y}}
\def\Nphi{N_{\phi}}
\def\Nalp{N_{\alpha}}
\def\gammatol{\gamma^{\text{tol}}}
\def\th{\ensuremath{\theta}}
\def\dth{\ensuremath{\dot{\theta}}}
\def\ddth{\ensuremath{\ddot{\theta}}}

\def\Cm{C_{\text{M}}}
\def\dCm{\dot{C}_{\text{M}}}

\newcommand{\epar}[1]{\ensuremath{e_{#1}}}
\newcommand{\bdn}{\mathcal{D}}
\newlength{\figwidth}
\newlength{\figwidthf}
\newlength{\figwidtha}
\newlength{\figwidthb}
\newlength{\figwidthc}
\newlength{\figwidthd}
\newlength{\figwidthe}
\let\originaleqref=\eqref
\renewcommand*{\eqref}[1]{Eq.~\originaleqref{#1}}

\graphicspath{{figs/}}

\title{Encoding nonlinear and unsteady aerodynamics of limit cycle oscillations using nonlinear sparse Bayesian learning
}

\author{
  Rimple Sandhu 
\thanks{\textit{Currently at National Renewable Energy Laboratory, USA}}\\
  Department of Civil and Environmental Engineering\\
  Carleton University \\
  Ottawa, ON, Canada \\
   \And
  Brandon Robinson\\
  Department of Civil and Environmental Engineering\\
  Carleton University \\
  Ottawa, ON, Canada\\
   \And
  Mohammad Khalil
\thanks{\textit{Sandia National Laboratories is a multimission laboratory managed and operated by National Technology and Engineering Solutions of Sandia, LLC., a wholly owned subsidiary of Honeywell International, Inc., for the U.S. Department of Energy’s National Nuclear Security Administration under contract DE-NA-0003525.}} \\ 
  Quantitative Modeling \& Analysis Department\\
  Sandia National Laboratories \\
  Livermore, CA, United States\\
   \And
  Chris L. Pettit\\
  Aerospace Engineering Department \\
  US Naval Academy\\
  Annapolis, MD, United States\\
   \And
  Dominique Poirel \\
  Department of Mechanical and Aerospace Engineering\\
  Royal Military College of Canada \\
  Kingston, ON, Canada \\
   \And
  Abhijit Sarkar \\
  Department of Civil and Environmental Engineering\\
  Carleton University \\
  Ottawa, ON, Canada\\
}

\begin{document}
\maketitle
\begin{abstract}
This paper investigates the applicability of a recently-proposed nonlinear sparse Bayesian learning (NSBL) algorithm to identify and estimate the complex aerodynamics of limit cycle oscillations. NSBL provides a semi-analytical framework for determining the data-optimal sparse model nested within a (potentially) over-parameterized model. This is particularly relevant to nonlinear dynamical systems where modelling approaches involve the use of physics-based and data-driven components. In such cases, the data-driven components, where analytical descriptions of the physical processes are not readily available, are often prone to overfitting, meaning that the empirical aspects of these models will often involve the calibration of an unnecessarily large number of parameters. While it may be possible to fit the data well, this can become an issue when using these models for predictions in regimes that are different from those where the data was recorded. In view of this, it is desirable to not only calibrate the model parameters, but also to identify the optimal compromise between data-fit and model complexity. In this paper, this is achieved for an aeroelastic system where the structural dynamics are well-known and described by a differential equation model, coupled with a semi-empirical aerodynamic model for laminar separation flutter resulting in low-amplitude limit cycle oscillations. For the purpose of illustrating the benefit of the algorithm, in this paper, we use synthetic data to demonstrate the ability of the algorithm to correctly identify the optimal model and model parameters, given a known data-generating model. The synthetic data are generated from a forward simulation of a known differential equation model with parameters selected so as to mimic the dynamics observed in wind-tunnel experiments.
\end{abstract}


\section{Introduction}
In this paper we demonstrate the applicability of the recently proposed nonlinear sparse Bayesian learning (NSBL) algorithm \cite{Sandhu2020, Sandhu2021} to a single degree of freedom (SDOF) aeroelastic oscillator that is undergoing low amplitude limit cycle oscillations. This experimental setup has been studied extensively through experimentation \cite{Poirel2008,Poirel2013}, numerical modelling of laminar separation flutter using high-fidelity large eddy simulations (LES) \cite{Poirel2010} and unsteady Reynolds averaged Navier-Stokes (URANS) model \cite{Poirel2011}. Furthermore, the wind tunnel experiments have provided a reliable test-bed for developing Bayesian techniques for system identification and model selection for nonlinear dynamical systems \cite{Sandhu2014,Sandhu2016,Bisaillon2022,Sandhu2017}. The work in \cite{Sandhu2014,Sandhu2016,Bisaillon2022} use standard methods of evidence-based Bayesian model selection, which allows for the systematic comparison of a set of candidate models with varying degrees of complexity. The model evidence as a criterion for model selection ensures a balance of favouring models with superior average data-fit, while penalizing models that are overly complex and thus prone to overfitting \cite{Muto2008}. In this context, model complexity is quantified by the KL-divergence of the parameter posterior probability density function (pdf) from the parameter prior pdf. For parameters where there exists little prior knowledge, it is typical to assign non-informative priors, however the width of the distribution used for the non-informative prior will influence the optimal complexity of the model. The issue of sensitivity to prior width is addressed in \cite{Sandhu2017}, whereby the problem is reposed as a sparse learning problem. Rather than non-informative priors, parameters with little prior information are assigned Gaussian automatic relevance determination (ARD) priors. The precision (inverse of the variance) of these ARD priors are determined through evidence optimization. In this re-framing of the inference problem, the optimal model is still quantified as such based on the model evidence. In contrast to the previous approach, rather than proposing an entire set of nested candidate models to determine the optimal model complexity, the current paper approaches the problem as an automatic discovery of the optimal sparse model nested within a single (potentially) over-parameterized model. Herein lies an additional benefit of approaching the problem as a sparse learning task; it is only necessary to obtain the parameter posterior for a single model, whereas standard methods require the calibration of each model in the candidate in order to then obtain an estimate of the model evidence. The shortcoming of this approach lies in the fact that the optimization process involves the use of Markov Chain Monte Carlo (MCMC) sampling at each iteration. This is addressed in the current NSBL framework, which removes the use of MCMC from within the optimization loop, resulting in significantly improved computational efficiency.

The NSBL framework presented here is an extension of the sparse Bayesian learning (SBL) also known as the relevance vector machine (RVM) algorithm \cite{Tipping2001,Faul01}. Both methods are motivated by the desire to avoid overfitting during Bayesian inversion. SBL/RVM and the similar Bayesian compressive sensing (BCS) algorithm \cite{Babacan10} provide analytical expressions for a sparse parameter posterior distribution owing to the analytical conveniences of the semi-conjugacy that exists between the Gaussian likelihood functions, and Gaussian ARD priors that are conditioned on hyperpriors that are Gamma distributions. The SBL methodology is extended to be applicable to nonlinear-in-parameter models and for non-Gaussian prior distributions, as these both commonly arise in engineering applications. We provide the minimum required mathematical details to understand the objectives of the algorithm and to provide a complete account of all terms shown in the equations used in this paper. For the full detailed derivation and additional details, we refer the reader to \cite{Sandhu2020, Sandhu2021}.


\section{Methodology: Nonlinear sparse Bayesian learning}
\label{sec:NSBL}
The NSBL methodology is applicable to general nonlinear mappings, $f: \bphi \mapsto \by$ where the model operator $f$ maps the unknown model parameter vector $\bphi \in \dR^{\Nphi}$ to the observable model output $\by \in \dR^{\Ny}$. In this specific application, $f$ represents the aeroelastic model, $\bphi$ are the deterministic system parameters and the stochastic parameters (relating to the model error), and $\by$ are the system output. Sensor measurements of the system output $\by$ at discrete points in time are denoted as $\bD$. The likelihood function $\pr(\bD|\bphi)$ can be computed for any $\bphi$, using the observations $\bD$, and these observation may be noisy, sparse, and incomplete measurements of the system state. The purpose of the algorithm is to obtain a data-optimal sparse representation of $\bphi$ using Bayesian inversion, while removing redundant parameters. 

NSBL operates within the following Bayesian framework; we seek the posterior distribution of the unknown model parameters $\bphi$ conditioned on the data $\bD$ and hyperparameters $\balpha$, 
\begin{equation}\label{eq:postpdf}
\pr(\bphi|\bD,\balpha) = \frac{\pr(\bD|\bphi)\pr(\bphi|\balpha)}{\pr(\bD|\balpha)} 
=\frac{\pr(\bD|\bphi)\pr(\bphi|\balpha)}{ \int \pr(\bD|\bphi)\pr(\bphi|\balpha) d\bphi}
\end{equation}
for given data and hyperparameters, the denominator, which represents the model evidence (or marginal likelihood or type-II likelihood), is just a normalization constant. The parameter prior $\pr(\bphi | \balpha)$ is also conditional on the hyperparameter. Though the objective is not to perform full hierarchical Bayesian inference, we nevertheless define a prior on $\pr(\balpha)$ (which is notably absent in the expression above); this hyperparameter prior (or hyperprior) becomes relevant during the optimization of $\balpha$. 

\begin{figure}[!ht]
\centering
\includegraphics[width=\textwidth]{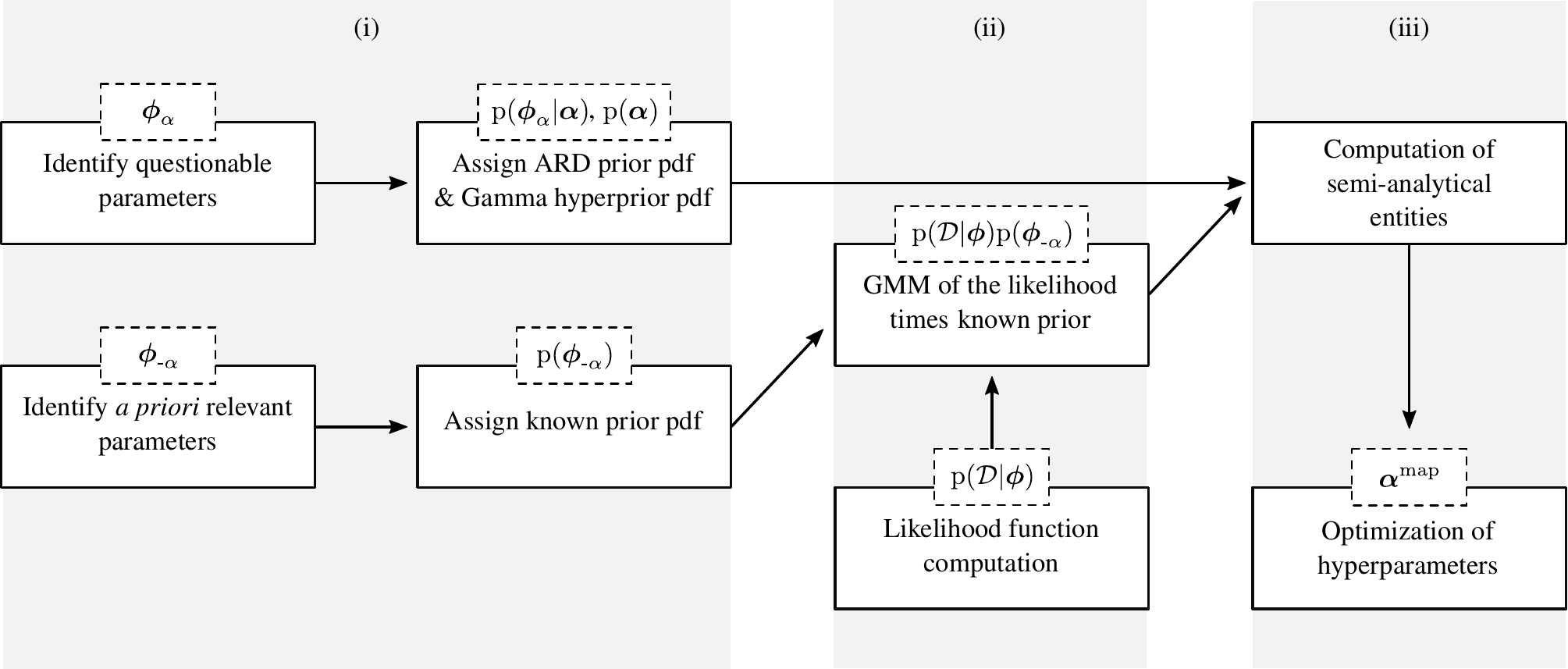}
\caption{Summary of the main steps involved in the NSBL algorithm.}
\label{fig:summary}
\end{figure}

The following sections outline the three principal tasks involved in the NSBL framework, as depicted in Figure~\ref{fig:summary}. Namely:

\begin{enumerate}[label=(\roman*)]
\item in~\cref{sec:hybrid_prior} we discuss the assignment of a \textit{hybrid prior}, wherein we distinguish between \textit{a priori} relevant parameters and questionable parameters, assigning known priors and ARD priors, respectively,
\item in~\cref{sec:gmm}, we detail the incorporation of data and the physics-based model through the construction of a Gaussian mixture model (GMM) over samples generated from the product of the likelihood function and the known prior, and 
\item in~\cref{sec:opt} we discuss the optimization of the hyperparameters. The derivation of various semi-analytical entities that enable the NSBL methodology is outlined in~\ref{sec:semi_analytical}. 
\end{enumerate} 

\subsection{Hybrid prior pdf}
\label{sec:hybrid_prior}
The model parameter vector $\bphi$ is first decomposed as $\bphi = \{\bphia,\bphima \}$, distringuishing between the set of parameters that are known to be relevant \textit{a priori}, denoted as $\bphia \in \dR^{\Nalp} $, and the parameters that the modeller has deemed to be questionable, denoted $\bphima \in \dR^{\Nphi-\Nalp}$. This classification as \textit{questionable} encompasses any parameter for which little or no prior knowledge exists, where a non-informative prior with large support would usually be used. The vector of questionable parameters are the set of parameters among which we will induce sparsity, as a subset of these parameters may be redundant. The mechanism for inducing sparsity follows SBL, where $\bphia$ is assigned a Gaussian ARD prior $\pr(\bphia|\balpha) = \cN(\bphia|\bZO,\bA^{-1})$. This prior is a normal distribution, whose mean vector is an $\Nalp \times 1$ zero vector, with a covariance matrix of $\bA^{-1}$, where $\bA$ is the precision matrix. Following SBL~\cite{Tipping01}, prior independence of the questionable parameters $\bphia$ is assumed, hence, the precision matrix is diagonal, $\bA = \text{diag}(\balpha)$. Furthermore, each parameter $\phi_i \in \bphia$ has a unique variable precision $\alpha_i$, such that we can write $\pr(\phi_i|\alpha_i) = \cN(\phi_i|0,\alpha_i^{-1})$. The hyperparameter $\alpha_i$ dictates the prior precision of parameter $\phi_i$; where low precision (or high variance) reduces to a non-informative prior, and conversely, a high precision (or low variance) results in an informative prior with a mean of zero. In the limit where the precision tends to infinity, the ARD prior becomes a Dirac delta function centered at zero, effectively pruning the parameter. Hence, the motivation behind NSBL  is that optimally selecting $\balpha$, can allow us to discover the model having the optimal complexity given the available data. The optimization criteria and methodology are presented later in~\cref{sec:opt}.

The joint prior pdf of $\bphi$ is summarized as \cite{Sandhu2021}
\begin{equation}
\pr(\bphi|\balpha) = \pr(\bphima) \pr(\bphia|\balpha)  =   \pr(\bphima) \cN(\bphia|\bZO,\bA^{-1}) . \label{eq:priorphi}  
\end{equation}
This hybrid prior pdf enables sparse learning of questionable parameters in $\bphia$ through the use of an ARD prior $\pr(\bphia|\balpha)$ while incorporating prior knowledge about parameters $\bphima$ through an informative prior $\pr(\bphima)$. The ARD prior is a conditional Gaussian distribution, whose precision depends on the hyperparameter $\balpha$. The marginal hyperprior pdf $\pr(\alpha_i)$ is chosen to be a Gamma distribution. Given the assumption of prior independence, the joint hyperprior $\pr(\balpha)$ is written as
\begin{equation} \label{eq:alphaprior}
\pr(\balpha) = \prod_{i=1}^{\Nalp} \pr(\alpha_i) = \prod_{i=1}^{\Nalp} \text{Gamma}(\alpha_i|r_i,s_i) = \prod_{i=1}^{\Nalp} \frac{s_i^{r_i}}{\Gamma(r_i)} \alpha_i^{r_i-1} e^{-s_i\alpha_i} ,
\end{equation}
where $\text{Gamma}(\alpha_i|r_i,s_i)$ denotes a univariate Gamma distribution parameterized by shape parameter $r_i>0$ and rate parameter $s_i>0$. The use of a Gamma distribution as the hyperprior allows us to enforce the requirement that the precision parameters $\balpha$ be positive. Furthermore, for specific combinations of shape and rate parameters, $r_i$ and $s_i$, the Gamma function can assume many forms of informative and non-informative priors. For instance, using values of $s_i \approx 1$ and $r_i \approx 0$ gives a flat prior over $\alpha_i$, or values of $s_i \approx 0$ and $r_i \approx 0$ gives Jeffreys prior, which is a flat over $\log\alpha_i$.  In fact, for reasons discussed in later sections, the NSBL algorithm operates on the natural logarithm of the hyperparameters rather than the hyperparameters directly. For this reason, we choose to use Jeffreys prior for the numerical results section. 

Using a univariate transformation of random variables~\cite{Papoulis2002}, the hyperprior in \eqref{eq:alphaprior} becomes 
\begin{equation}
\pr(\log\balpha) = \prod_{i=1}^{\Nalp} \pr(\log \alpha_i) =  \prod_{i=1}^{\Nalp}\frac{\pr(\alpha_i)}{\left\vert\frac{d }{d \alpha_i} \log \alpha_i \right\vert}  =    \prod_{i=1}^{\Nalp} \frac{s_i^{r_i}}{\Gamma(r_i)} \alpha_i^{r_i} e^{-s_i\alpha_i} . \label{eq:logalphaprior}
\end{equation}



\subsection{Gaussian mixture-model approximation}\label{sec:gmm}
After defining the joint parameter prior distribution as in Eq.~(\ref{eq:priorphi}), we substitute the resulting expression into the conditional posterior distribution from Eq.~(\ref{eq:postpdf}), yielding \cite{Sandhu2021}
\begin{equation}\label{eq:hybridpostpdf}
\pr(\bphi|\bD,\balpha) = \ \frac{\pr(\bD|\bphi)\pr(\bphima) \cN(\bphia|\bZO,\bA^{-1})}{\pr(\bD|\balpha)}.
\end{equation}

Given that the ARD priors are normally distributed, NSBL constructs a GMM approximation of the remaining terms in the numerator, $\pr(\bD|\bphi)\pr(\bphima)$, which  enables the derivation of semi-analytical expressions for many entities of interest. Obtaining expressions for the model evidence and objective function (\ref{sec:evid_postpdf}), the parameter posterior (\ref{sec:parpostpdf}), and the gradient and Hessian of the objective function (\ref{sec:grad}) makes the optimization of the hyperparameters analytically tractable. Moreover,the use of a GMM enables the preservation of non-Gaussianity in both the likelihood function and the known prior. The construction involves the estimation of kernel parameters $\ak$, $\bmuk$, and $\bSigmak$ , 

\begin{equation} \label{eq:likapprox}
\pr(\bD|\bphi)\pr(\bphima) \approx  \sum_{k=1}^{K} \ak \cN(\bphi|\bmuk,\bSigmak) ,
\end{equation}
where $K$, $\ak$, and $\cN(\bphi|\bmuk,\bSigmak)$ are the total number of kernels, the kernel coefficient ($\ak>$0), and a Gaussian pdf with mean vector  $\bmuk \in \dR^{\Nphi}$ and covariance matrix $\bSigmak \in \dR^{\Nphi \times \Nphi}$ \cite{Sandhu2021}. For a Gaussian likelihood function and a Gaussian known prior, this reduces to the case of SBL or RVM \cite{Tipping01}, and a single kernel is sufficient. Otherwise, except in the case of a Laplace approximation~\cite{Bishop2006} of the likelihood function times the known prior, multiple kernels will generally be required. The construction of the GMM typically involves the use of MCMC in order to generate samples from the arbitrary distribution, which requires repeated function evaluations for different samples of the unknown parameter vector. The model itself operates as a black-box, thus, the analytical form of the model does not need to be known; the model is only needed in order to compute the likelihood function. Once samples have been generated from the posterior distribution, the estimation of the kernel parameters in Eq. (\ref{eq:likapprox}) can be performed numerically using methods such as kernel density estimation (KDE) or expectation maximization (EM) \cite{Murphy2022}. Since the construction of the GMM involves numerous forward solves of the model, this step is the most computationally demanding component of the algorithm. Notably, the GMM itself is independent of the hyperparameters, thus this process only needs to be performed once at the onset, and does not need to be repeated during the optimization of the hyperparameters.

\subsection{Sparse learning optimization problem}
\label{sec:opt}
The critical step in the NSBL algorithm is the optimization of the hyperparameter, $\balpha$. Within a hierarchical Bayesian framework, we seek a point estimate for the hyperparameters, rather than obtaining posterior estimates thereof. As in SBL, we perform type-II maximum likelihood, seeking the values $\balpha$, which maximize the hypeperparameter posterior, 

\begin{equation} \label{eq:alphapost}
\pr(\balpha|\bD) = \frac{\pr(\bD|\balpha)\pr(\balpha)}{\pr(\bD)} \propto \pr(\bD|\balpha)\pr(\balpha).
\end{equation}
for a fixed set of data $\bD$, the denominator $\pr(\bD)$ is a normalization constant that is analytically intractable in general. Thus for optimization, we consider only the numerator which is the product of the model evidence $\pr(\bD|\balpha)$ and the hyperprior $\pr(\balpha)$ . The type-II maximum a posteriori (MAP-II) estimate $\balphaMAP$ can therefore be posed as \cite{Murphy2022}
\begin{equation} \label{eq:alphamap}
\balphaMAP = \argmax_{\balpha} \{ \pr(\balpha|\bD)\} = \argmax_{\balpha} \{ \pr(\bD|\balpha)\pr(\balpha)\} .
\end{equation}
Since the natural logarithm is a strictly increasing function, we re-write the optimization problem in terms of  $ \log \pr(\balpha|\bD)$, which facilitates the derivation of the semi-analytical expressions of the gradient and Hessian.
Furthermore, as noted above, the optimization is performed with respect to $\log\balpha$ instead of $\balpha$. This helps to account for the potentially widespread difference in the scales of the prior precision between relevant parameters (whose $\alphaiMAP$ are finite) and irrelevant parameters (whose $\alphaiMAP$ tend to infinity). This has the added benefit of automatically enforcing the positivity constraint of $\balpha$ throughout the optimization procedure. Restating~\eqref{eq:alphamap} as the optimization of the log-evidence with respect to the log of the hyperparameters $\balpha$, the objective function $\cL(\log\balpha)$ becomes

\begin{align}
\log\balphaMAP &= \argmax_{\log\balpha} \{\cL(\log\balpha) \}  \nonumber \\
&=  \argmax_{\log\balpha} \{ \log \pr(\log \balpha|\bD)\} \nonumber \\
& =  \argmax_{\log\balpha} \{\log \hat{\pr}(\bD|\log\balpha) + \log \pr(\log\balpha)\} \label{eq:logalpha} .
\end{align}

The objective function is the sum of the estimate of the model evidence and the hyperprior (where constant terms that are independent of $\alpha_i$ are discarded). As the model evidence itself will be analytically intractable in general, it is replaced by the estimate $\hat{\pr}(\bD|\log\balpha)$ in \eqref{eq:evidfinal} that is available in terms of the $K$ kernels of the GMM.  Substituting \eqref{eq:logalphaprior} for the hyperprior gives

\begin{align} \label{eq:cL}
\cL(\log\balpha) & = \log \hat{\pr}(\bD|\log\balpha) + \sum_{i=1}^{\Nalp} \left(r_i \log \alpha_i - s_i \alpha_i\right) .
\end{align}

In this form, it becomes clear that when Jeffreys prior ($s_i \approx 0$, $r_i \approx 0$) is used, the objective function reduces to the log-evidence, resulting in the common procedure known as the type-II maximum likelihood estimate or Emperical Bayes method~\cite{Murphy2022}, which may lead to a non-convex optimization problem~\cite{Bishop2006}. The numerical examples that follow will illustrate the possibility of non-unique optima in the objective function. The possibility of multiple optima can be addressed using a multi-start optimization algorithm, where the optimization routine is initiated from an array of different coordinates of $\log \balpha$ in an effort to discover all local optima in order to determine the global optima. Sequential estimates of $\{\log\balpha_j\}$ are obtained starting from the initial iterate $\{\log\balpha_0\}$, according to \cite{Sandhu2021}

\begin{equation}
\log\balpha_{j+1} = \log\balpha_j + \beta_j\bp_j ,
\end{equation}
where $\beta_j$ is the step-length~\cite{Nocedal06}. The optimization itself is expedited by the ability to derive the gradient vector $\bj(\log\balpha_i)$ and Hessian matrix $\bH(\log\balpha_i)$ of the objective function from \eqref{eq:cL}, as outlined in \eqref{eq:Jacobian} and \eqref{eq:Hessian}, respectively. This permits the convenient calculation of $\bp_j$ as the solution to

\begin{equation}
 \bH(\log\balpha_j)\bp_j =-\bj(\log\balpha_j)\label{eq:NewtonIter}.
\end{equation}
The specific method for determining $\bp_j$ (e.g., modified Newton method, trust-region Newton method) must consider that the Hessian is not guaranteed to be a positive definite matrix~\cite{Nocedal06}.



The optimization of the objective function (the log hyperparameter posterior) with respect to $\log \balpha$, provides the MAP estimate $\log\balphaMAP$. As discussed previously, a large value of $\log \alphaMAP_i$ indicates high prior precision of parameter $\phi_i$ and implies it is redundant. However, it is difficult to quantify \textit{high precision} directly as this is highly parameter dependent. To remove the scale-dependence, we leverage another concept from SBL~\cite{Tipping01}, wherein a relevance indicator is defined for each parameter $\phi_i$, according to $\gamma_i = 1 - \alpha_i P_{ii}$, where, $P_{ii}$ is the $i$th diagonal entry of the posterior covariance matrix and $\alpha_i$ is the $i$th diagonal entry in the prior precision matrix, $\bA$. In NSBL, it is possible to define a similar metric, however, as with the semi-analytical entities described previously, a relevance indicator will be defined for each of the $K$ Gaussian kernels. This can be re-written as a ratio of prior to posterior precision \cite{Sandhu2021},  

\begin{equation} \label{eq:gammaik}
\gammaik = 1- \frac{\alpha_i}{(\Piik)^{-1}} \in [0,1] ,
\end{equation}
hence, the relevance indicator provides a normalized metric on the scale of 0 to 1, where a value of 0 indicates irrelevance, while a value of 1 implies relevance. The intuition behind this conclusion is that for a parameter where the ratio of prior-to-posterior precision is close to unity, the posterior precision is dictated by the prior, rather than the likelihood, suggesting that the parameter does not learn from the data. The converse holds as well, where parameters that do learn from the data will tend to have higher posterior precision compared to the prior precision, hence the expression for the relevance indicator will approach a value of 1. Since we perform this one a kernel-by-kernel basis, to summarize the relevance indicators for a given parameter across all kernels in the GMM, we propose the use of a root-mean-square value of the relevance indicator \cite{Sandhu2021}

\begin{equation} \label{eq:gammaisum}
\gammaisum = \left(\frac{1}{K} \sum_{k=1}^{K} (\gammaik)^2\right)^{1/2} =  \left(\frac{1}{K} \sum_{k=1}^{K} \left(1 - \alpha_i \Piik\right)^2 \right)^{1/2} .
\end{equation}

%
%

\section{Application to aeroelastic oscillator}
In this section, we demonstrate the performance of NSBL for identifying sparsity in the unknown parameters of nonlinear stochastic differential equations. We consider a single degree-of-freedom pitching airfoil undergoing low-amplitude limit cycle oscillations. The structural model in \eqref{eq:struct_vdp} consists of a typical mass-spring-damper system augmented by a cubic stiffness and a model for dry friction. The structural dynamics are coupled with a semi-empirical aerodynamic model in \eqref{eq:aero_vdp}, which we will refer to as an unsteady generalized Duffing-Van der Pol model \cite{Poirel2010}. The aerodynamic moment coefficient ($C_M$) modelled as a first-order ODE, which allows for unsteadiness in the aerodynamics, driven by the parameter $B$. The model features a polynomial expansion of the pitch angle ($\theta$) and pitch rate ($\dot{\theta}$) to attempt to capture nonlinearity in the aerodynamic moment coefficient. Finally, it contains a model error term to capture the discrepancy between the modelled physics and the true, but ultimately unknown, aerodynamic loads. The resulting system of coupled ODEs are written in terms of non-dimensional time ($\tau$) as \cite{Sandhu2016}

\begin{subequations}
\begin{align}
I \ddot{\theta} + C \dot{\theta} + K \theta + C_{nl} \text{sign}(\dot{\theta})+ K_{nl} \theta^3  &= \frac{1}{2} \rho U^2 c s C_M \label{eq:struct_vdp} \\
\frac{\dot{C}_M}{B} + C_M &= a_0 + a_1 \theta + a_2 \dot{\theta} + a_3 \theta^2 + a_4 \theta \dot{\theta} + a_5 \dot{\theta}^2 + \hdots +  \frac{c_6}{B} \ddth  + \sigma \xi(\tau)\label{eq:aero_vdp}
\end{align}\label{eq:vdp}
\end{subequations}
where the structural parameters $I$, $C$,  $C_{nl}$ $K$ and $K_{nl}$, and aerodynamic parameters $\rho$, $U$, $c$, $s$, and $c_6$ are known precisely, and coefficients $B$, $\sigma$, and aerodynamic coefficients $a_i$ are to be estimated.

\begin{figure}[!ht]
\centering
\includegraphics[width=\figwidth]{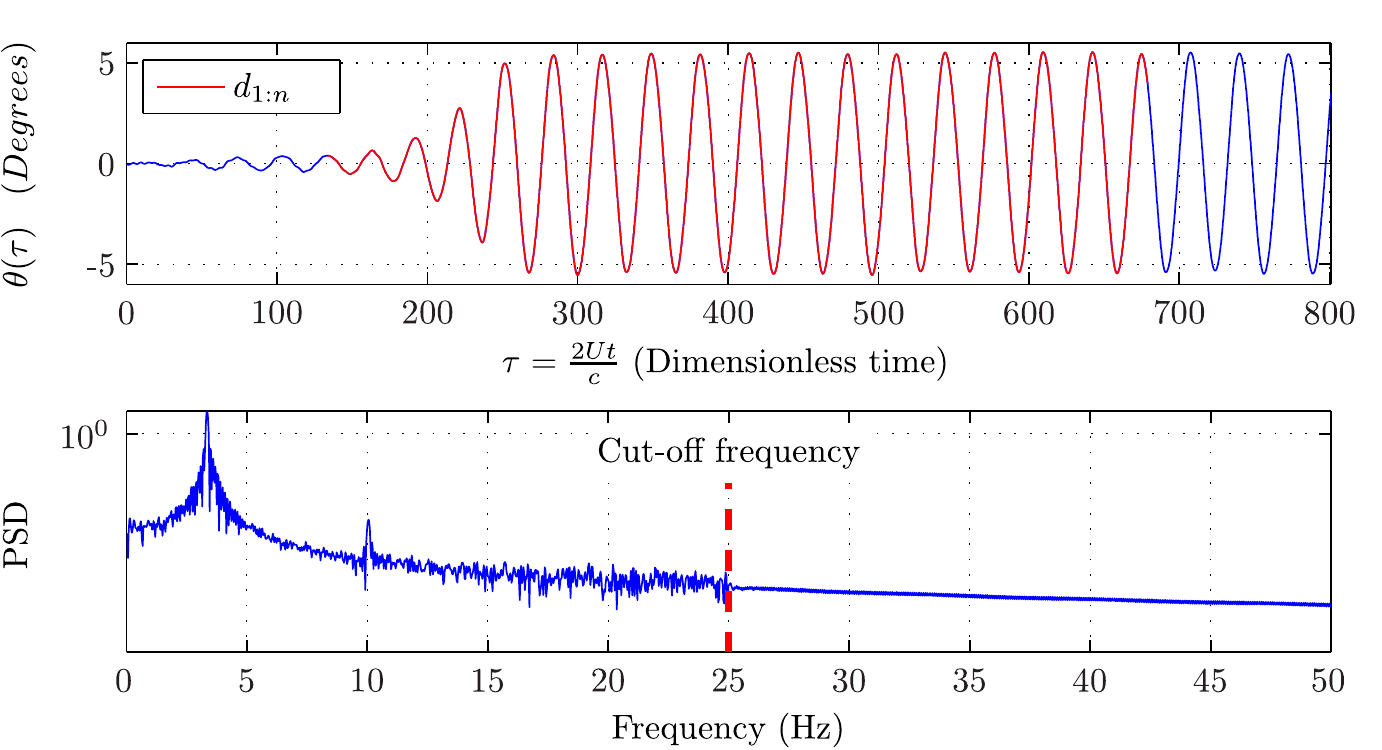}
\caption{Synthetic pitch measurements and its PSD. The data used for ARD computations is shown in red. (Reprinted from Sandhu et al. \cite{Sandhu2017}, with permission from Elsevier).}
\label{fig:num_filt}
\end{figure}

In the current study, synthetic LCO data is used, in order to validate the use of NSBL in discovering the data-optimal sparse model. In this scenario, the data-generating model and the candidate model have the same underlying analytical form, though the candidate model may contain extra parameters that were not present (i.e. were equal to 0) when generating the data. Hence the algorithm should ideally recover the parameters used in the data-generating model, and identify any parameters that are not used to generate the data as irrelevant. The synthetic LCO from which the measurements of the pitch are generated have similar frequency and amplitude as are observed in the wind-tunnel experimental setup. Synthetic measurements are generated by recording the pitch deflection (representing only a partial measurement of the state). Measurements are recorded at a frequency of 1000Hz (which is consistent with the temporal density of observations available experimentally). These measurements are then corrupted by additive Gaussian noise with a noise strength that again mimics the noise levels in the wind-tunnel experiments. As in the physical experiments, a low-pass filter is used to remove contributions with a frequency above 25Hz (see the bottom pannel of Figure \ref{fig:num_filt}). In practice, the sensors capture contributions from the wind-tunnel motor at 30Hz. The dominant frequency of the system is 3.25Hz, so many super-harmonics remain below the cut-off frequency.

The data-generating model is shown in Eq. (\ref{eq:vdp}). The structural model from Eq. (\ref{eq:struct_vdp}) is rewritten in a standardized form, replacing the system parameters by coefficients $c_1, \hdots, c_6$. The aerodynamic model is also rewritten, retaining only a subset of the terms listed in Eq. (\ref{eq:aero_vdp}), and replacing the aerodynamic parameters by coefficients $e_1, \hdots, e_4$. The terms retained are limited to: (i) linear stiffness $e_1 \theta$, (ii) linear damping $e_2 \dot{\theta}$, (iii) nonlinear Duffing-type stiffness $e_3 \theta^3$, and (iv) Van der Pol-type nonlinear damping term $e_4 \theta ^2 \dot\theta$. Furthermore, unsteadiness is introduced through the coefficient $B$, and a random forcing is introduced through $\sigma \xi(\tau)$, where $\xi(\tau)$ is a white noise process

\begin{subequations} \label{eq:dgODEnsbl}
\begin{align}
\ddth& = c_1 \text{sign}(\dth) + c_2 \th + c_3 \Cm + c_4 \dth + c_5 \th^3 , \label{eq:sODEnsbl} \\
\frac{\dCm}{B} + \Cm& = \epar{1} \th + \epar{2} \dth + \epar{3} \th^3 + \epar{4} \th^2 \dth + \frac{c_6}{B} \ddth  + \sigma \xi(\tau). \label{eq:aODE}
\end{align}
\end{subequations}
The structural parameters $c_1, \hdots, c_6$, from \eqref{eq:sODEnsbl} and the aeroedynamic coefficients $B, e_1, \hdots, e_4$ and the strength of the random forcing $\sigma$ in \eqref{eq:aODE} are summarized in Table \ref{tab:synthetic_data}. The observations, $\bdn$ are shown in Figure~\ref{fig:num_filt}. Next, we exploit the NSBL algorithm from \Cref{sec:NSBL} to demonstrate the algorithm's ability to correctly identify the relevance or irrelevance of parameters in one-, two-, and four-dimensional sparse learning exercises. 

{
\renewcommand{\arraystretch}{1.6}
\begin{table}[!ht]
\small
\centering
\begin{tabular}{p{0.3\textwidth}|p{0.3\textwidth}}
\hline
Structural parameters & Aerodynamic parameters \\
\hline
 $c_1=-6.875 \times 10^{-5}$ & $B= 2.000 \times 10^{-1}  $\\
 $c_2=-2.038 \times 10^{-2}$ & $e_1= -1.250  $\\
 $c_3=-3.819 \times 10^{-2}$ & $e_2= -1.000$ \\
 $c_4=-7.275 \times 10^{-3}$ & $e_3= 1.000\times 10^{2}$ \\
 $c_5=1.824 \times 10^{-1}$ & $e_4= - 5.000\times 10^{2}$ \\
 $c_6= -2.507 \times 10^{-1} $ & $\sigma= 2.000 \times 10^{-3}$ \\
\hline 
\end{tabular}
\caption{Structural and aerodynamic model parameters used to generate the synthetic data.} 
\label{tab:synthetic_data} 
\end{table}
}

\subsection{Unidimensional sparse learning}
\label{sec:unidim_num}
We first consider a unidimensional sparse learning problem, meaning the questionable parameter vector $\bphia$ and the associated hyperparameter vector $\balpha$ are both scalar entities. Given the measurements $\bdn$, the proposed model for $\Cm$ is chosen to be the same as the data-generating model shown in \eqref{eq:aODE}. The structural equation of motion in \eqref{eq:sODEnsbl} is assumed to be known for the sake of inverse modelling. The cubic aerodynamic stiffness parameter $e_3$ (or term $e_3 \th^3$) in \eqref{eq:aODE} is treated as a questionable parameter, whose relevance to the LCO aerodynamics needs to be determined using the NSBL algorithm. As per \Cref{sec:hybrid_prior}, $\bphi$ is decomposed into the questionable parameter $\bphia = \{e_3\}$ and the \textit{a priori} relevant parameter vector $\bphima$ = $\{B$, $e_1$, $e_2$, $e_4$, $\sigma\}$. The prior pdf $\pr(\bphima)$ is known, while $e_3$ is assigned an ARD prior $\cN(e_3|0,\alpha^{-1})$. \Cref{tab:ess1D_setup} summarizes this unidimensional sparse learning setup, including the known prior pdf of $\bphima$. Note that $\cLN(.|r,s)$ represents a log-normal distribution with median at $r$ and coefficient of variation of $s$; and $\cU(.|e,f)$ represents a uniform distribution with lower bound $e$, and upper bound $f$.

{
\renewcommand{\arraystretch}{1.6}
\begin{table}[!ht]
\footnotesize
\centering
\begin{tabular}{c|c}
\hline
Aerodynamic model & $\displaystyle \frac{\dCm}{B} + \Cm = e_1 \th + e_2 \dth + e_3 \th^3 + e_4 \th^2 \dth + \frac{c_6}{B} \ddth  + \sigma \xi(\tau)$ \\
 $\bphi$ decomposition & $\bphia$ = $\{e_3\}$ ,\;\; $\bphima$ = $\{B$, $e_1$, $e_2$, $e_4$, $\sigma\}$ \\ \hline  
ARD prior, $\pr(\bphia|\alpha)$ & $\cN(e_3|0,\alpha^{-1})$  \\
Known prior, $\pr(\bphima)$ &  $ \cLN(B|0.2, 0.5) \cU(e_1| {-2},0) \cU(e_2| {-2},0) \cU(e_4|{-600},0) \cLN(\sigma |0.002,0.5) $  \\ \hline 
\end{tabular}
\caption{Unidimensional sparse learning setup where parameter $e_3$ is treated as questionable.}
\label{tab:ess1D_setup} 
\end{table}
}

The likelihood function $\pr(\bdn|\bphi)$ is computed using the extended Kalman filter (EKF) as outlined in \cite{Sandhu2014,bisaillon2015} to propagate the joint state pdf through the nonlinear model. Given the known prior pdf $\pr(\bphima)$ from \Cref{tab:ess1D_setup}, the unnormalized pdf $\pr(\bdn|\bphi)\pr(\bphima)$ is sampled using the MCMC sampler. A total of 5000 stationary (i.e. post burn-in) MCMC samples are generated from the six-dimensional pdf $\pr(\bdn|\bphi)\pr(\bphima)$. These 5000 samples possess some correlation due to the Markovian nature of MCMC sampling. In an effort to alleviate the effect of correlation, every tenth sample is extracted to produce 500 independent and identically distributed (iid) samples from $\pr(\bdn|\bphi)\pr(\bphima)$. These 500 iid samples are then decomposed to into 10 sets of 50 iid samples each. These iid sets are used to investigate the finite sample properties of NSBL algorithm. Note that this sampling of $\pr(\bdn|\bphi)\pr(\bphima)$ is only required to be performed once since both $\pr(\bdn|\bphi)$ and $\pr(\bphima)$ remain unchanged during sparse learning. Moreover, generating 5000 stationary samples from a six-dimensional pdf is computationally cheap using MCMC algorithms such as delayed rejection adaptive metropolis (DRAM) \cite{ haario2006dram} and transitional MCMC \cite{ching2007transitional}  

\begin{figure}[!ht]
\centering
\includegraphics[width=\figwidthf]{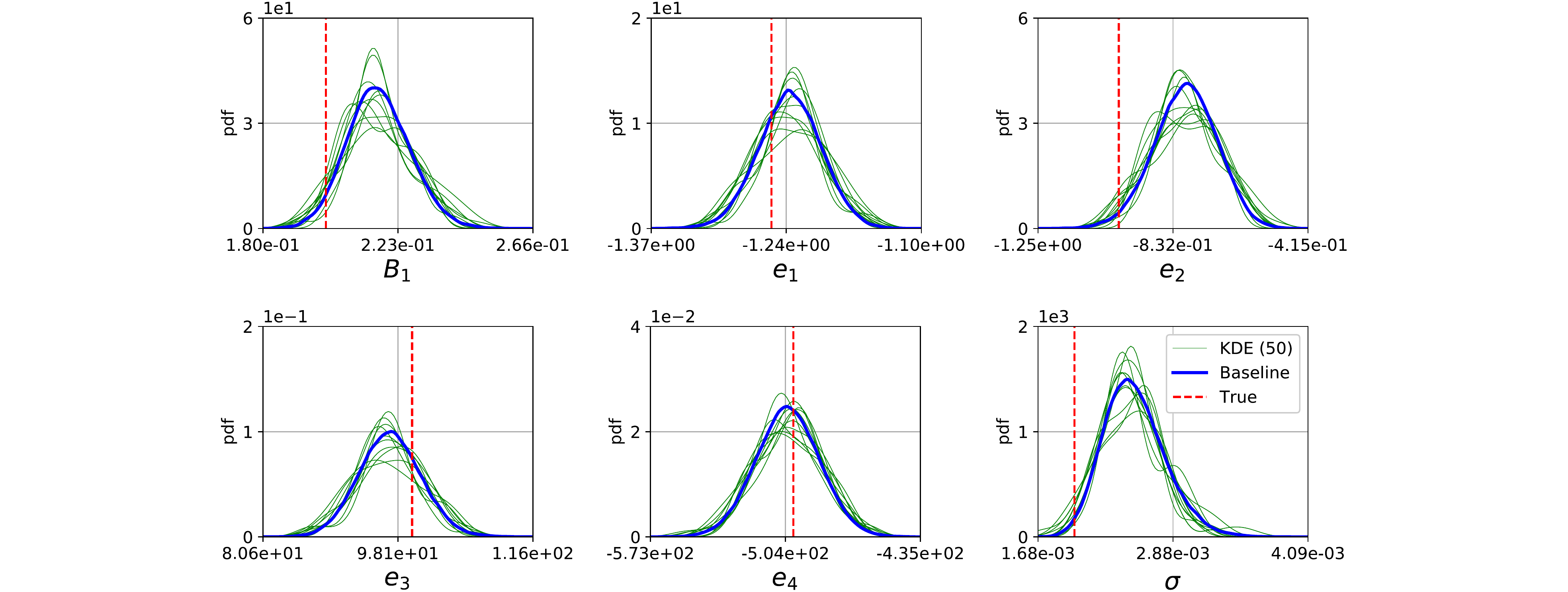}
\caption{Marginal parameter pdf obtained using 10 instances of 50 iid samples from $\pr(\bdn|\bphi)\pr(\bphima)$. The baseline (converged) pdf is obtained by using all of the 500 iid samples.}
\label{fig:ess1D_mpdf}
\end{figure}

Next, a multivariate KDE approximation is built for $\pr(\bdn|\bphi)\pr(\bphima)$ using each of the 10 sets of 50 iid samples. When using a Gaussian kernel, the KDE approximation resembles the kernel-based approximation in \eqref{eq:likapprox}. Henceforth, the number of kernels $K$ is the same as the number of samples, which is 50 for the current case. The coefficient $\ak$ are all equal to one since the Gaussian kernels in KDE are all equally weighted. The mean vector $\bmuk$ for each kernel is equal to the corresponding iid sample values (vector). The covariance matrix $\bSigmak$ is computed automatically using Scott's rule for estimating KDE bandwidth~\cite{Scott92} and is the same for all 50 kernels. \Cref{fig:ess1D_mpdf} shows the marginal parameter pdf pertaining to each of the 10 instances of KDE approximation of $\pr(\bdn|\bphi)\pr(\bphima)$. The marginal pdfs show a reasonable variation across multiple instances considering only 50 iid samples were used.

\begin{figure}[!ht]
\centering
\includegraphics[width=\figwidthf]{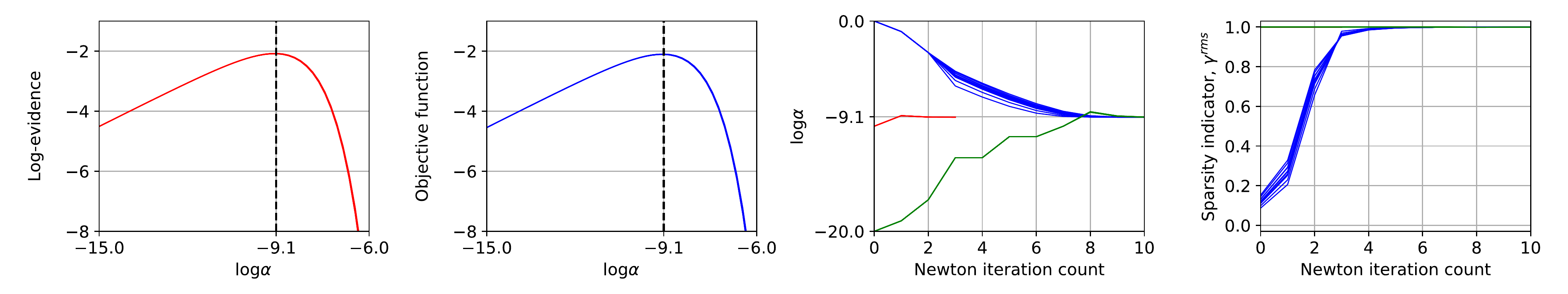}
\caption{NSBL results for the relevance determination of an \textit{a priori} relevant parameter $e_3$.}\label{fig:ess}
\end{figure}

\Cref{fig:ess} shows the objective function $\cL(\log\alpha)$ and the log-evidence $\log\hat{\pr}(\bdn|\log\alpha)$ for varying $\log\alpha$ value, pertaining to each instance of the KDE approximation. There is little-to-no variability observed in the objective function and log-evidence across the different instances of KDE. The objective function is computed using hyperprior parameters of $\log r_i$ = $\log s_i$ = -6.0 in \eqref{eq:cL}. The similarity between the two leftmost panels implies the negligible effect of hyperprior in $\cL(\log\alpha)$ for the range of $\log\alpha$ values shown. Also, the variation in $\cL(\log\alpha)$ across different KDE instances is much less than that for the marginal pdfs shown in \Cref{fig:ess1D_mpdf}. This behaviour of $\cL(\log\alpha)$ indicates that the relevance property of a parameter is more robust to sampling variability than the KDE approximation of $\pr(\bdn|\bphi)\pr(\bphima)$.

The optimization of the objective function ($\cL(\log\alpha)$) and the associated value of the relevance indicator ($\gamma^{\text{rms}}$) are shown as a function of Newton iteration count in the two right-most panels of \Cref{fig:ess}, respectively. The $\log \alpha$ iterates during the multistart Newton iteration, and the corresponding value of the relevance indicator $\gammaisum$ from \eqref{eq:gammaisum} are shown. The $\log \alpha$ iterates and $\gammaisum$ are shown for each of the 10 KDE instances of $\pr(\bdn|\bphi)\pr(\bphima)$. The Newton iteration is initiated from three $\log \alpha$ values of -20.0, -10.0 and 0.0. The analytical solution for the gradient and Hessian of $\cL(\log \alpha)$ employed in the Newton iteration was first validated using a finite-difference scheme (results not reported here for brevity). All the multistart Newton iterations converge to a unique optimum of $\alphaMAP$ = -9.1. The relevance indicator $\gammasum$ converges to a value of one, implying the strong relevance of parameter $e_3$ to the LCO aerodynamics. This relevance of $e_3$ is insensitive to the choice of tolerance $\gammatol$ as the optimal relevance indicator is very close to one. The posterior pdf of $\bphi$ can be obtained as per \ref{sec:parpostpdf} using the optimal ARD prior $\cN(e_3|0,\text{exp}(9.1))$. Since this optimal prior has large variance, the posterior parameter pdf obtained from NSBL is the same as that obtained by sampling $\pr(\bdn|\bphi)\pr(\bphima)$ (the exact pdf shown in \Cref{fig:ess1D_mpdf}). In other words, the optimal ARD prior did not bias the posterior pdf of $e_3$. 

Next, consider the unidimensional sparse learning setup in \Cref{tab:ness1D_setup} where the aerodynamic model is augmented by a higher-order nonlinear stiffness term $e_5 \theta^5$. Here, parameter $e_5$ is treated as questionable. Given that this term was not used to generate the data, we should expect the $e_5$ parameter to be deemed irrelevant by NSBL. 

{
\renewcommand{\arraystretch}{1.5}
\begin{table}[!ht]
\footnotesize
\centering
\begin{tabular}{c|c}
\hline 
Aerodynamic model &  $\displaystyle \frac{\dCm}{B} + \Cm = e_1 \th + e_2 \dth + e_3 \th^3 + e_4 \th^2 \dth + e_5\th^5 + \frac{c_6}{B} \ddth  + \sigma \xi(\tau)$ \\
 $\bphi$ decomposition & $\bphia$ = $\{e_5\}$ ,\;\; $\bphima$ = $\{B$, $e_1$, $e_2$, $e_3$, $e_4$, $\sigma\}$ \\ \hline  
ARD prior, $\pr(\bphia|\balpha)$ & $\cN(e_5|0,\alpha^{-1})$ \\
Known prior, $\pr(\bphima)$ &  $ \cLN(B|0.2, 0.5)\, \cU(e_1| {-2},0) \, \cU(e_2| {-2},0) \, \cU(e_3| {-250},250)\cU(e_4|{-600},0) \, \cLN(\sigma |0.002,0.5) $ \\ \hline 
\end{tabular}
\caption{Unidimensional sparse learning setup where parameter $e_5$ is treated as questionable.}
\label{tab:ness1D_setup} 
\end{table}
}

\begin{figure}[!ht]
\centering
\includegraphics[width=\figwidthf]{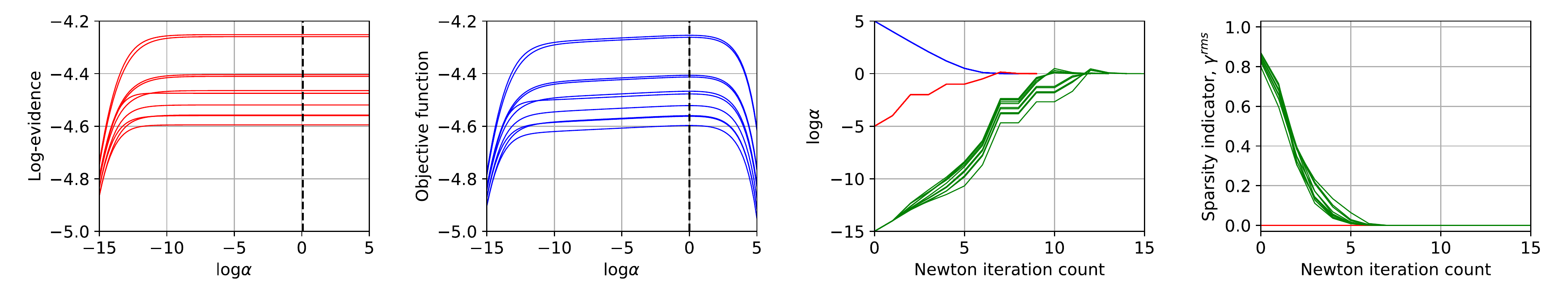}
\caption{NSBL results for the relevance determination of an \textit{a priori} irrelevant parameter $e_5$.}
\label{fig:ness}
\end{figure}

The two left-most pannels of \Cref{fig:ness} show the log-evidence $\log \hat{\pr}(\bdn|\log\alpha)$ and the objective function $\cL(\log\alpha)$ using the hyperprior parameters of $\log r_i$ = $\log s_i$ = -6. Unlike the previous case, both these entities possess large sampling variability across multiple KDE instances of $\pr(\bdn|\bphi)\pr(\bphima)$. However, the relevance of $e_5$ is determined by the optimum of $\cL(\log\alpha)$, which remains same across multiple KDE instances. This finite sample property of $\cL(\log\alpha)$ reiterates the notion that relevance is well-determined using limited iid samples from $\pr(\bdn|\bphi)\pr(\bphima)$, even when the KDE instances possess large sampling variability. Also notice that, unlike log-evidence, the objective function $\cL(\log\alpha)$ is free from flat regions for large $\log\alpha$ values. The Hessian matrix is singular in flat regions. Therefore, the presence of flat regions in log-evidence makes it unsuitable for the application of Newton's method. The absence of flat regions in $\cL(\log\alpha)$ is due to the influence of hyperprior $\pr(\alpha)$ for large $\log\alpha$ values. This ensures $\cL(\log\alpha)$ is amenable to optimization by Newton's method wherein the Hessian is non-singular at all times. The shape and rate parameters of the Gamma hyperprior must have positive values, hence to assign Jeffreys prior, they are given values that approach zero from above. If these values are sufficiently small (here we use $\log r_i$ = $\log s_i$ = -6.0), they should have a negligible effect of the relevance of parameters, but from \eqref{eq:cL}, it is understood that they will help regularize the sparse learning optimization problem for large values of $\balpha$.

\begin{figure}[!ht]
\centering
\includegraphics[width=\figwidthf]{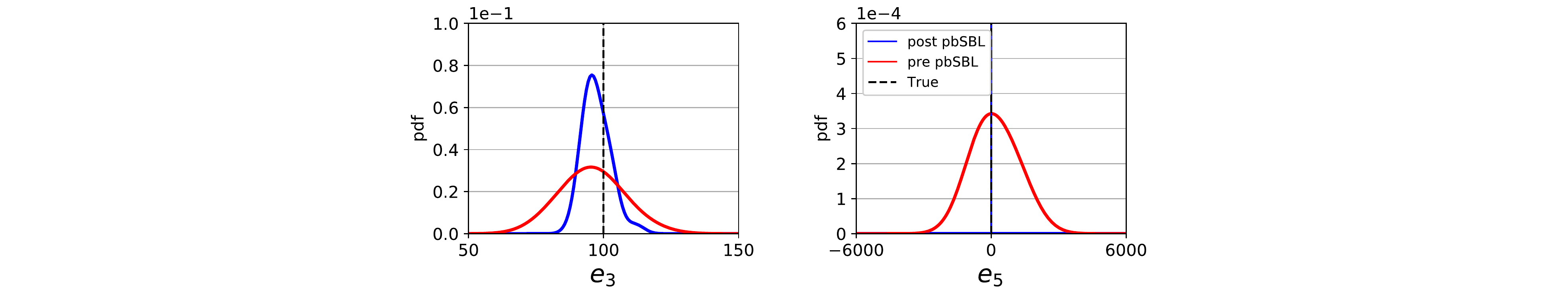}
\caption{NSBL results for the relevance determination of an \textit{a priori} irrelevant parameter $e_5$.}
\label{fig:ness-pdf}
\end{figure}

The two rightmost plots of Figure \ref{fig:ness} show the $\log\alpha$ iterates and the RMS relevance indicator $\gammasum$; when initiating at $\log\alpha$ values of -15.0, -5.0 and 5.0. The multistart Newton iteration converges to a unique optimum of $\alphaMAP$ = 0 and $\gammasum$ = 0 for all multistart iterations and for all KDE instances of $\pr(\bdn|\bphi)\pr(\bphima)$. Note that since the objective function is relatively flat close to the optimum, the estimated optimum can vary with a varying function tolerance set for terminating the Newton iteration. However, the key entity to monitor is the relevance indicator. Even with a small variation in the optimum hyperparameter value, the relevance indicator converges to the value of zero. According to \eqref{eq:gammaisum}, a $\gammasum$ value of zero implies irrelevance since the posterior pdf is entirely dictated by the Dirac-delta ARD prior centered at zero. This change in the posterior pdf of $e_5$ following NSBL is shown in the right pannel of \Cref{fig:ness-pdf}.

\Cref{fig:ness-pdf} shows the posterior pdf of both nonlinear stiffness parameters $e_3$ and $e_5$ before and after the inclusion of optimal ARD prior $\cN(e_5|0,\text{exp}(0.0))$. Notice that the posterior pdf of $e_3$ following the removal of $e_5$ has a lower uncertainty and is in close agreement with the true value. This behaviour can be explained by realizing that $e_3$ and $e_5$ both quantify the nonlinear aerodynamic stiffness. When both the parameters are present in the model, overfitting occurs. When $e_5$ is removed, overfitting is remedied, and $e_3$ is estimated with an increased accuracy using $\bdn$. This numerical exercise demonstrates the benefit of sparse learning in preventing overfitting and enabling robust probabilistic predictions outside the regime of measurements. The two unidimensional cases studied here instills confidence in NSBL as a practical tool to perform sparse learning among physics-based models. Next, we study two-dimensional sparse learning cases to gain a more in-depth understanding of NSBL.

\subsection{Two-dimensional sparse learning}
\label{sec:twodim_num}
In this section, we consider two-dimensional sparse learning cases where the number of questionable parameters is two (i.e. $\Nalp$ = 2). The hyperparameter vector is $\balpha$ = $\{\alpha_1, \alpha_2\}$. NSBL is executed with the same settings as \Cref{sec:unidim_num}. Hence, the likelihood function $\pr(\bdn|\bphi)$ is computed using EKF, followed by the generation of 500 iid samples from $\pr(\bdn|\bphi)\pr(\bphima)$ using the DRAM MCMC algorithm. NSBL is executed for each of the 10 KDE instances of $\pr(\bdn|\bphi)\pr(\bphima)$ containing 50 iid samples each. Also, the hyperprior parameters are chosen as $\log r_i$ = $\log s_i$ = -6.0 for all the cases reported in this section. We consider the three possible combinations of relevant and irrelevant parameters: where the questionable parameters are both relevant, where one is relevant and the other is irrelevant, and where both are irrelevant.

\begin{figure}[!ht]
\centering
\includegraphics[width=\figwidthf]{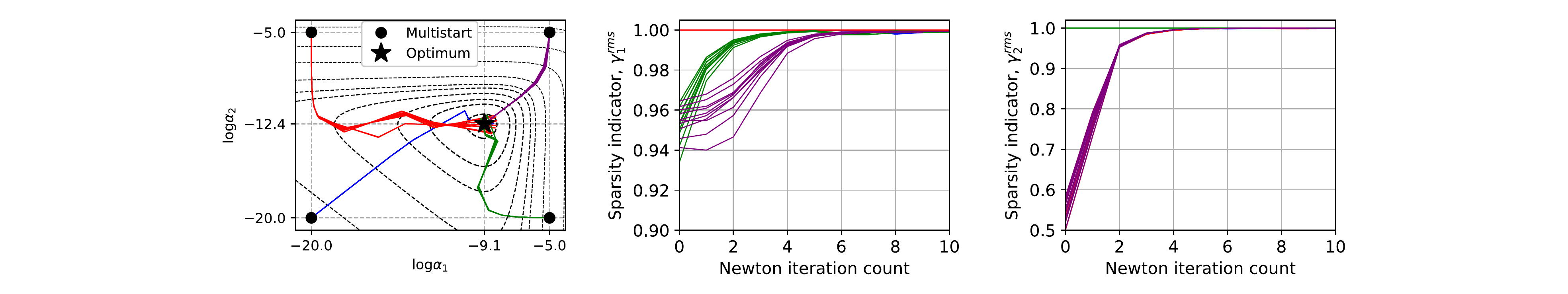}
\caption{NSBL results for the relevance determination of parameter $e_3$ and $e_4$.}
\label{fig:essess}
\end{figure}

We first consider the Bayesian setup in \Cref{tab:essess2D_setup} where the proposed model is same as the data-generating model from \eqref{eq:aODE}, and parameters $e_3$ and $e_4$ are treated as questionable. \Cref{fig:essess} shows the $\log\balpha$ iterates during Newton iteration initiated from \{-20, -20\}, \{-5, -20\}, \{-20, -5\} and \{-5, -5\}, for each of the 10 KDE instances. Notice that all of the multistart Newton iterations converges to a unique optimum of $\log\balphaMAP$ = \{-9.1, -12.4\}. Also, convergence to a unique optimum for varying KDE instances of $\pr(\bdn|\bphi)\pr(\bphima)$ demonstrates that the curvature of objective function $\cL(\log\balpha)$ remains fairly insensitive to sampling variability. The two rightmost panels of \Cref{fig:essess} show the variation in RMS relevance indicator $\gammasum$ for parameter $e_3$ and $e_4$, respectively. Both the relevance indicators converge to one, implying the relevance of both the parameters to the LCO physics. NSBL converges in less than 10 Newton iterations, demonstrating the power of the Hessian informed optimizer.

{
\renewcommand{\arraystretch}{1.6}
\begin{table}[!ht]
\footnotesize
\centering
\begin{tabular}{c|c}
\hline
Aerodynamic model & $\displaystyle \frac{\dCm}{B} + \Cm = e_1 \th + e_2 \dth + e_3 \th^3 + e_4 \th^2 \dth + \frac{c_6}{B} \ddth  + \sigma \xi(\tau)$ \\
 $\bphi$ decomposition & $\bphia$ = $\{e_3, e_4\}$ ,\;\; $\bphima$ = $\{B$, $e_1$, $e_2$, $\sigma\}$ \\ \hline  
ARD prior, $\pr(\bphia|\alpha)$ & $\cN(e_3|0,\alpha_1^{-1})$ $\cN(e_4|0,\alpha_2^{-1})$ \\
Known prior, $\pr(\bphima)$ &  $ \cLN(B|0.2, 0.5) \cU(e_1| {-2},0) \cU(e_2| {-2},0)\cLN(\sigma |0.002,0.5) $  \\ \hline 
\end{tabular}
\caption{Two-dimensional sparse learning setup where parameters $e_3$ and $e_4$ are treated as questionable.}
\label{tab:essess2D_setup} 
\end{table}
}

\begin{figure}[!ht]
\centering
\includegraphics[width=\figwidthf]{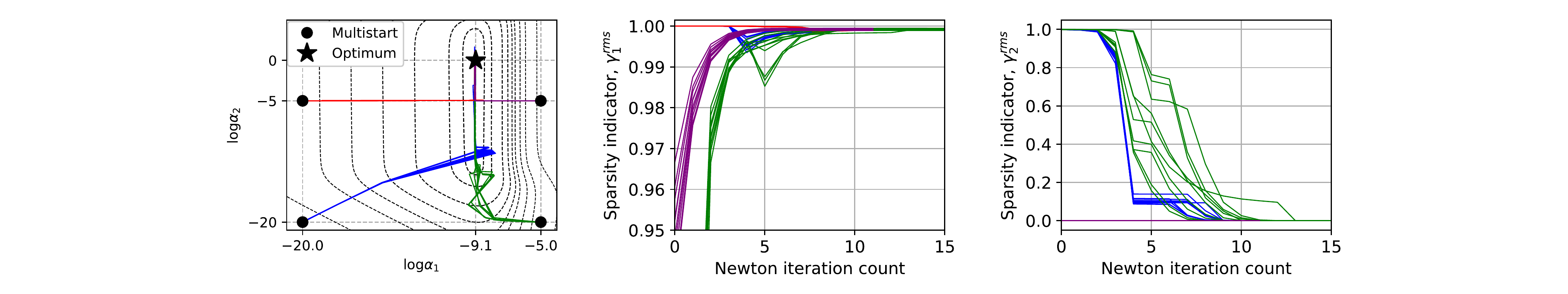}
\caption{NSBL results for the relevance determination of parameter $e_3$ and $e_5$.}
\label{fig:essness}
\end{figure}

Next, consider the Bayesian setup in \Cref{tab:essness2D_setup} where parameters $e_3$ and $e_5$ are treated as questionable. The data-generating model in \eqref{eq:aODE} contains the cubic aerodynamic stiffness coefficient $e_3$, while the fifth-order aerodynamic stiffness coefficient $e_5$ is absent. \Cref{fig:essness} shows the $\log\balpha$ iterates pertaining to starting $\log\balpha$ values of \{-20, -20\}, \{-5, -20\}, \{-20, -5\} and \{-5, -5\}. The multistart Newton iteration converges to a unique optimum of $\log\balphaMAP$ = \{-9.1, 0.0\}. Recall that these are the two values obtained when these parameters were studied independently in \cref{sec:unidim_num}. The two right-most panels of \Cref{fig:essness} shows the RMS relevance indicators for parameter $e_3$ and $e_5$, respectively. Parameter $e_3$ is determined to be relevant since $\gammasum_1$ converges to exactly one. On the contrary, parameter $e_5$ is determined to be irrelevant as $\gammasum_2$ converges to zero. These conclusion are true for all multistart locations and for all KDE instances of $\pr(\bdn|\bphi)\pr(\bphima)$.

{\renewcommand{\arraystretch}{1.6}
\begin{table}[!ht]
\footnotesize \centering
\begin{tabular}{c|c}
\hline
Aerodynamic model & $\displaystyle \frac{\dCm}{B} + \Cm = e_1 \th + e_2 \dth + e_3 \th^3 + e_4 \th^2 \dth + e_5 \th^5 + \frac{c_6}{B} \ddth  + \sigma \xi(\tau)$ \\
 $\bphi$ decomposition & $\bphia$ = $\{e_3, e_5\}$ ,\;\; $\bphima$ = $\{B$, $e_1$, $e_2$, $e_4$, $\sigma\}$ \\ \hline  
ARD prior, $\pr(\bphia|\alpha)$ & $\cN(e_3|0,\alpha_1^{-1})$ $\cN(e_5|0,\alpha_2^{-1})$ \\
Known prior, $\pr(\bphima)$ &  $ \cLN(B|0.2, 0.5) \cU(e_1| {-2},0) \cU(e_2| {-2},0)\cU(e_4|{-600},0) \cLN(\sigma |0.002,0.5) $  \\ \hline
\end{tabular}
\caption{Two-dimensional sparse learning setup where parameters $e_3$ and $e_5$ are treated as questionable.}
\label{tab:essness2D_setup} 
\end{table}}

\begin{figure}[!ht]
\centering
\includegraphics[width=\figwidthf]{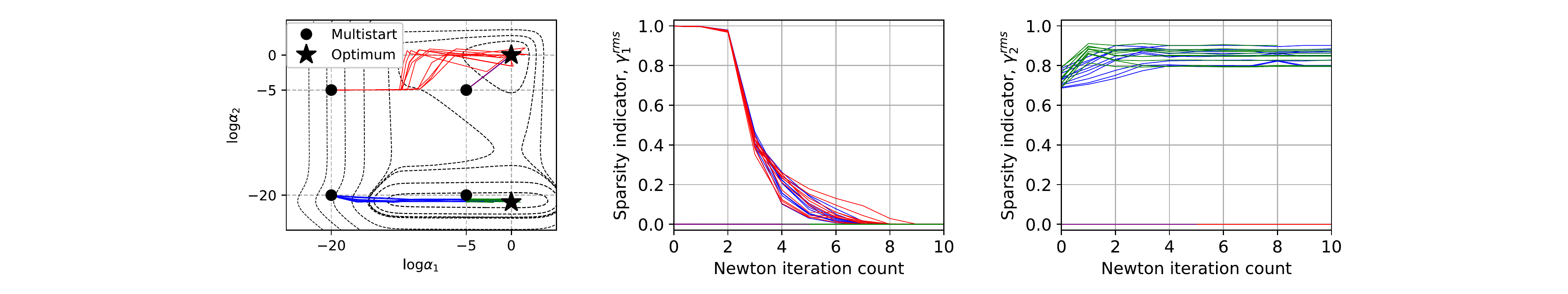}
\caption{NSBL results for the relevance determination of parameter $e_5$ and $e_6$.}
\label{fig:nessness}
\end{figure}

Next, consider the Bayesian setup in \Cref{tab:nessness2D_setup}, where parameters $e_5$ and $e_6$ are treated as questionable. Both these parameters were absent from the data-generating model in \eqref{eq:aODE}. \Cref{fig:nessness} shows the $\log\balpha$ iterates for the multistart Newton iteration beginning from \{-20, -20\}, \{-5, -20\}, \{-20, -5\} and \{-5, -5\}. The Newton iterations converge to two different optimums of \{0.0, -20.5\} and \{0.0, 0.0\}. Based on the $\cL(\log\balpha)$ values at these optimums, \{0.0, -20.5\} was found to the global optimum (for the range of $\alpha_1$ and $\alpha_2$ considered here). This ability of the multistart Newton iteration to capture multiple optima was further validated using the case of bimodal likelihood functions.

{\renewcommand{\arraystretch}{1.6}
\begin{table}[!ht]
\footnotesize \centering
\begin{tabular}{c|c}
\hline
Aerodynamic model & $\displaystyle \frac{\dCm}{B} + \Cm = e_1 \th + e_2 \dth + e_3 \th^3 + e_4 \th^2 \dth + e_5 \th^5 +  e_6 \th^4 \dth + \frac{c_6}{B} \ddth  + \sigma \xi(\tau)$ \\
 $\bphi$ decomposition & $\bphia$ = $\{e_5, e_6\}$ ,\;\; $\bphima$ = $\{B$, $e_1$, $e_2$, $e_3$, $e_4$, $\sigma\}$ \\ \hline  
ARD prior, $\pr(\bphia|\alpha)$ & $\cN(e_5|0,\alpha_1^{-1})$ $\cN(e_6|0,\alpha_2^{-1})$ \\
Known prior, $\pr(\bphima)$ &  $ \cLN(B|0.2, 50) \cU(e_1| {-2},0) \cU(e_2| {-2},0)\cU(e_3|{-250},{250})  \cU(e_4|-1\times 10^4,0)  \cLN(\sigma |0.002,0.5) $  \\ \hline
\end{tabular}
\caption{Two-dimensional sparse learning setup where parameters $e_5$ and $e_6$ are treated as questionable.}
\label{tab:nessness2D_setup} 
\end{table}}

\Cref{fig:nessness} shows the RMS relevance indicator for parameter $e_5$ and $e_6$ during multistart Newton iterations. Parameter $e_5$ is rendered irrelevant irrespective of the optimum as $\gammasum_1$ converges to zero for both the optima. The parameter $e_6$ requires a special attention as $\gammasum_2$ in \Cref{fig:nessness} converges to a value in the range [0.75,0.90] for the case of the global optimum.  As demonstrated previously in the unidimensional setting, $\gammaisum$ value converges to exactly one for relevant parameters. Also, based on our experience, the convergence of relevance indicator $\gammaisum$ to one is a necessary condition for a parameter to be deemed relevant. This condition implies that the posterior pdf of a relevant parameter should be largely dictated by the likelihood function and not the data-informed ARD prior. Based on this principle, the parameter $e_6$ should be deemed irrelevant. Alternatively, one could define a tolerance $\gammatol$ for determining the relevance of such parameters. A $\gammatol$ value of $0.5$ implies parameter $e_6$ as relevant, whereas a $\gammatol$ value of $0.9$ implies an irrelevant $e_6$. This choice of an appropriate $\gammatol$ imparts greater flexibility to the modeller, thereby allowing the sparse learning process to align with the modelling goals.

\begin{figure}[!ht]
\centering
\includegraphics[width=\figwidthf]{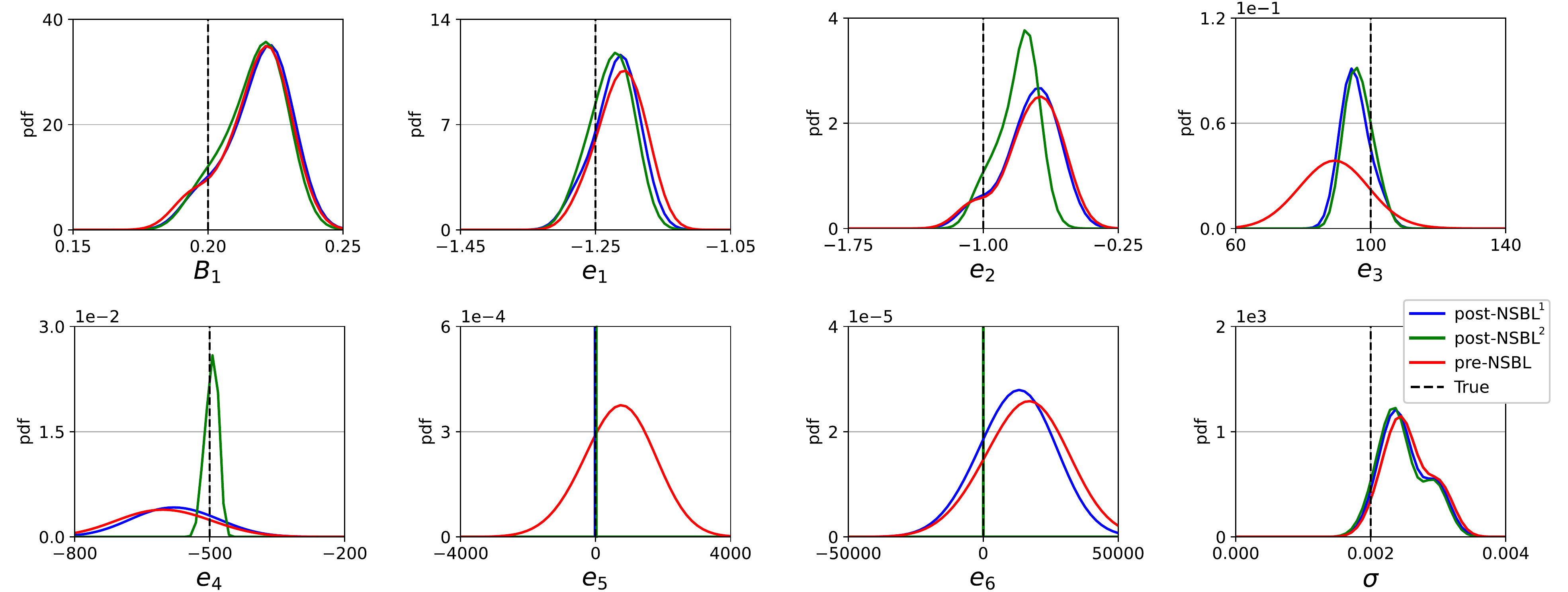}
\caption{Marginal posterior pdf of model parameters before and after sparse learning, for the case when $e_5$ and $e_6$ are deemed irrelevant.}
\label{fig:nessness2D_pdf}
\end{figure}

\Cref{fig:nessness2D_pdf} shows the marginal parameter pdf before and after the inclusion of the optimal ARD prior for questionable parameters $e_5$ and $e_6$ at both optima identified in \Cref{fig:nessness}. The pre-NSBL pdf, shown in blue, is obtained from the KDE representation of $\pr(\bdn|\bphi)\pr(\bphima)$ using 500 iid samples. The post-NSBL pdfs for the case where $e_5$ is irrelavant, but $e_6$ are relevant is shown in red. The case where both $e_5$ and $e_6$ are irrelevant are shown in green. These posterior pdfs are obtained using the analytical solution derived in \ref{sec:parpostpdf}. First, in red, where only $e_5$ is correctly identified as irrelevant,the posterior pdf of $e_5$ reduces to a Dirac-delta function centered at zero. This eliminates all of the uncertainty in the estimate for $e_5$, and also can be shown to reduce the uncertainty in the estimate for $e_3$, which is also a nonlinear stiffness parameter. Second, in green, where both $e_5$ and $e_6$ are correctly identified as irrelevant and are both reduced to Dirac-delta function centered at zero, there is additional reduction in uncertainty observed for both parameters $e_4$ and $e_2$, which are nonlinear and linear damping coefficients, respectively. The removal of these higher-order nonlinear terms, however, is shown here to have a minimal effect on the posterior estimate for $\sigma$. This behaviour of the marginal posterior pdf of $\sigma$ indicates that the removal of redundant parameters during NSBL does not affect the data-fit property of the model. Had NSBL identified an overly-simplistic model, the model error strength would be expected to increase. Notice the decrease in posterior uncertainty of parameters $e_3$ and $e_4$ following the removal of $e_5$ and $e_6$. This decrease is attributed to the remediation of overfitting by the sparse learning process.      \Cref{fig:nessness2D_pdf} shows the corresponding marginal posterior pdfs for the case when $e_5$ is deemed irrelevant and $e_6$ is deemed relevant.

The objective function $\cL(\log\balpha)$ was observed to be dictated by log-evidence for the range of $\balpha$ values shown in \Cref{fig:nessness}, meaning both log-evidence and $\cL(\log \balpha)$ are bimodal. To our best knowledge, this is the first reporting of a multimodal log-evidence for a physics-based Bayesian inverse modelling. This multimodality also demonstrates the issue of \textit{global identifiability} in the model (or hyperparameter) space, while the inverse problem may or may not be globally identifiable in the likelihood (or model parameter) space. This multimodality also warrants careful consideration of evidence-based Bayesian model comparison while dealing with nested models with closely-related nonlinear terms.

NSBL was also executed for cases with 100, 500, and 1000 iid samples per KDE instance of $\pr(\bdn|\bphi)\pr(\bphima)$. These results are not reported here since they were identical to those reported in this section but with even less variation in $\log\balpha$ iterates and relevance indicator $\gammasum$ across multiple instances. This conclusion was true for the three numerical cases studied in this section. 

\subsection{Four-dimensional sparse learning}
\label{sec:fourdim_num}
\begin{figure}[!ht]
\centering
\includegraphics[width=\figwidthf]{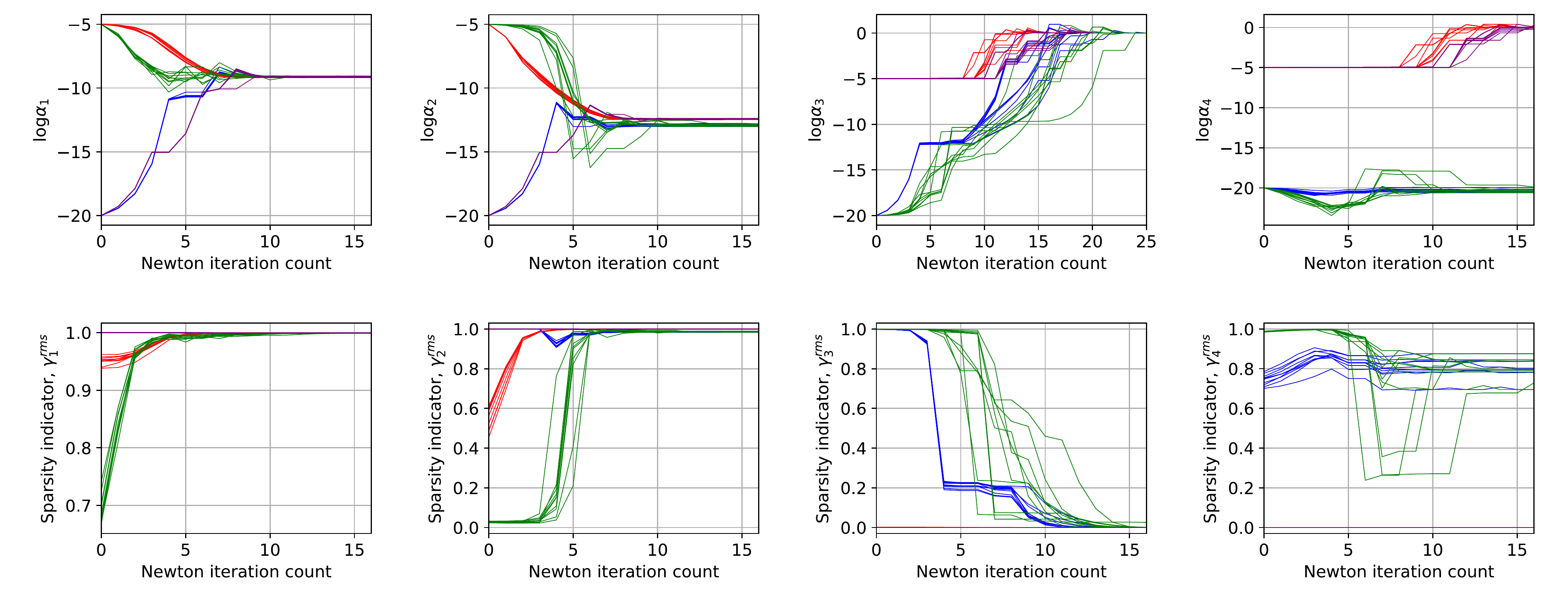}
\caption{NSBL results for the four-dimensional sparse learning.}
\label{fig:4D_results}
\end{figure}

\begin{figure}[!ht]
\centering
\includegraphics[width=\figwidthf]{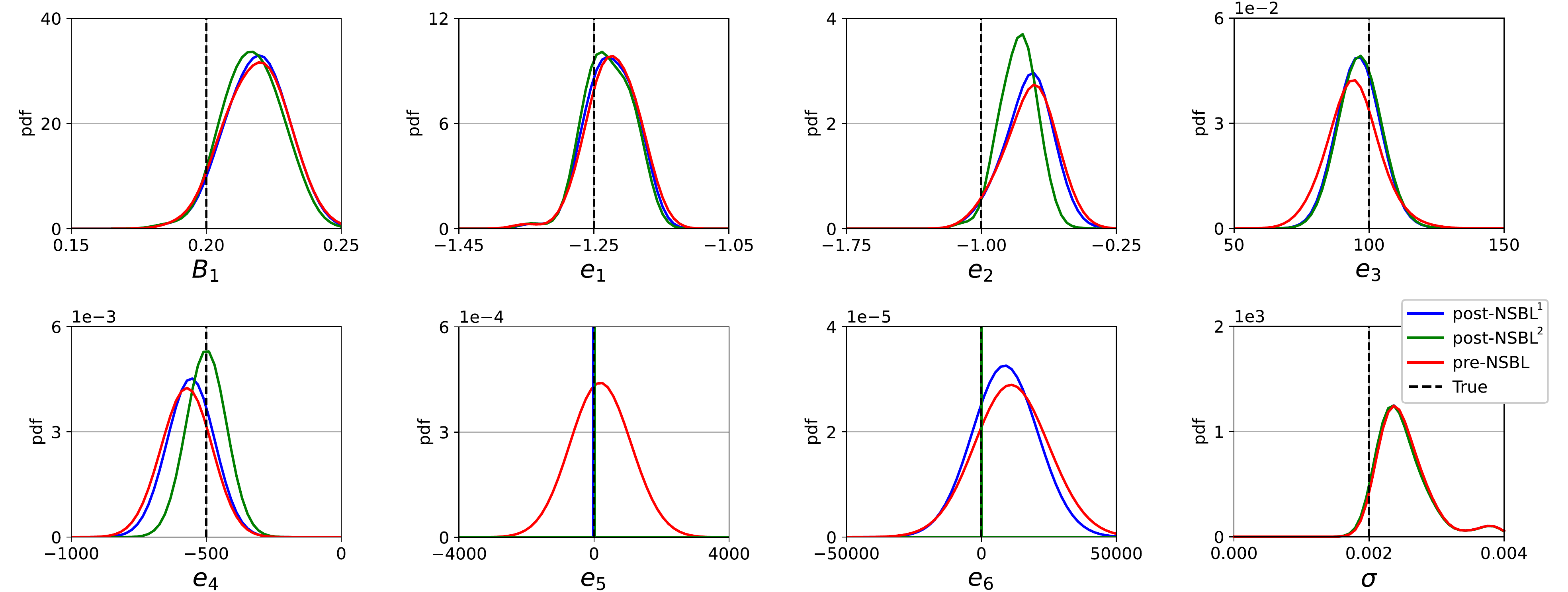}
\caption{Marginal posterior pdf of model parameters before the after the inclusion of optimal ARD prior.}
\label{fig:4D_pdf}
\end{figure}

Consider the four-dimensional sparse learning problem detailed in \Cref{tab:4D_setup}. Here we only consider a single case, where among four questionable parameters, two parameters are relevant, and two are irrelevant. NSBL is executed to determine the relevance of questionable parameters $e_3$, $e_4$, $e_5$ and $e_6$. The NSBL setup is the same as \Cref{sec:unidim_num}. \Cref{fig:4D_results} shows the $\log\balpha$ iterates and relevance indicator $\gammaisum$ for multistart Newton iterations beginning from \{-20, -20, -20, -20\}, \{-20, -20, -5, -5\}, \{-5,-5, -20, -20\} and \{-5, -5, -5, -5\}. Notice that all but $\log\alpha_4$ converges to a unique optimum. The $\log\alpha_4$ converges to two different optimums depending on the starting point. This type of behaviour has been previously reported in \Cref{fig:nessness}. Based on the $\cL(\log\balpha)$ values at these optimums, $\log\alphaMAP_4$ = 0.0 was found to the global optimum. This produces the solution $\log\balphaMAP$ = \{-9.1, -12.4, 0.0, 0.0\}, and the corresponding value of relevance indicator as \{1.0, 1.0, 0.0, 0.0\}. This optimal value of $\gammaisum$ indicates that parameter $e_3$ and $e_4$ are important, while parameter $e_5$ and $e_6$ are redundant. This conclusion is in agreement with the data-generating model in \eqref{eq:aODE} that was used to generate $\bdn$. \Cref{fig:4D_pdf} shows the marginal posterior pdf before and after the inclusion of the optimal ARD prior computed using NSBL. The pre-NSBL pdf is obtained using 500 iid samples from $\pr(\bdn|\bphi)\pr(\bphima)$, while the post-NSBL pdf is obtained using the kernel-based analytical solution derived in \ref{sec:parpostpdf}. Once again, in Figure \ref{fig:4D_pdf}, we show the pre-NSBL results in blue, the local optimum (where $e_5$ is relevant and $e_6$ is relevant) in red, and the global optimum (where both $e_5$ and $e_6$ are irrelevant) in green. Notice the decrease in uncertainty in relevant parameters following the removal of parameter $e_5$ and $e_6$. Note that the level of sparsity in physical models such as the LCO models is significantly lower than data-based models due to the presence of physics-based parameters. Significant benefits are realized in terms of decrease in uncertainty in the posterior pdf even with the removal of two parameters.

{\renewcommand{\arraystretch}{1.6}
\begin{table}[!ht]
\footnotesize \centering
\begin{tabular}{c|c}
\hline 
Aerodynamic model &  $\displaystyle \frac{\dCm}{B} + \Cm = e_1 \th + e_2 \dth + e_3 \th^3 + e_4 \th^2 \dth + e_5\th^5 + e_6 \th^4 \dth  + \frac{c_6}{B} \ddth  + \sigma \xi(\tau)$ \\
 $\bphi$ decomposition & $\bphia$ = $\{e_3, e_4, e_5, e_6\}$ ,\;\; $\bphima$ = $\{B$, $e_1$, $e_2$, $\sigma\}$ \\ \hline  
ARD prior, $\pr(\bphia|\balpha)$ & $\cN(e_3|0,\alpha_1^{-1})\cN(e_4|0,\alpha_2^{-1})\cN(e_5|0,\alpha_3^{-1})\cN(e_6|0,\alpha_4^{-1})$, \\
Known prior, $\pr(\bphima)$ &  $ \cLN(B|0.2, 0.5)\, \cU(e_1| {-2},0) \, \cU(e_2| {-2},0) \cLN(\sigma |0.002,0.5)$  \\ \hline 
\end{tabular}
\caption{Four-dimensional sparse learning setup.}
\label{tab:4D_setup} 
\end{table}}

\section{Conclusion}
\label{sec:conclusion}
With the use of synthetic data, we have demonstrated that NSBL is a promising approach to computationally efficient model selection for nonlinear stochastic dynamical systems. In all cases studied herein, the algorithm has shown its ability to recapture the data-generating model parameters and correctly identify the relevance of the parameters. Note that for this case study, the data is not particularly sparse temporally, nor is it particularly noisy, though, the data only represents a partial observation of the state. For the given data, even an overly complex model may not necessarily yield poor predictions of the response, nonetheless, it has been shown that the parameter estimates can be significantly improved by the removal of unnecessary parameters. 

The specific aspects of NSBL that make it particularly well suited to engineering systems is the ability to use non-Gaussian known priors, and the applicability to systems where there is a nonlinear mapping from the uncertain parameters to the observed outputs. In the aeroelastic example, the underlying physics and/or laws of statistics prevented certain parameters from assuming non-positive values, hence a Gaussian distributions would not be proper choices of prior. Likewise, the LCO being modelled is a nonlinear phenomenon, which naturally leads to a nonlinear-in-parameter model in the machine learning context. Both of these qualities are enabled by the use of a GMM approximation of the product of the likelihood function and the known prior. This GMM also enables the semi-analytical machinery that enables the efficient computation of estimates of many quantities of interest including the parameter posterior distribution, the model evidence, the objective function, Jacobian/ gradient vector and Hessian matrix.

Finally, we have illustrated the need for robust optimization scheme as there is the potential for multiple optima in the objective function, corresponding to different locally optimal values of the hyperparameter, corresponding to different classifications of the parameter as relevant versus irrelevant.

\bibliographystyle{unsrt}
\bibliography{my_library}

\appendix

\section{Semi-analytical calculation of Bayesian entities}\label{sec:semi_analytical}
\subsection{Model evidence}
\label{sec:evid_postpdf}
The model evidence was first introduced as the normalization factor in \eqref{eq:postpdf}.  Substituting the expression for the joint prior pdf in \eqref{eq:priorphi}, the model evidence can be rewritten as \cite{Sandhu2021}
\begin{equation} \label{eq:evid0}
\pr(\bD|\balpha) =\int \pr(\bD|\bphi)\pr(\bphi|\balpha) d\bphi = \int \pr(\bD|\bphi) \pr(\bphima) \pr(\bphia|\balpha) d\bphi .
\end{equation}

Now, substituting the GMM approximation of the product of the likelihood function and the known prior from \eqref{eq:likapprox}, and the expression for the ARD prior, $\pr(\bphi|\balpha)$ = $\cN(\bphia|\bZO,\bA^{-1})$, \eqref{eq:evid0} becomes

\begin{align}
\hat{\pr}(\bD|\balpha)&= \int \left\{\sum_{k=1}^{K} \ak \cN(\bphi|\bmuk,\bSigmak)\right\} \cN(\bphia|\bZO,\bA^{-1}) d\bphi \nonumber \\
& = \sum_{k=1}^{K} \ak \int \cN(\bphi|\bmuk,\bSigmak) \cN(\bphia|\bZO,\bA^{-1}) d\bphi. \label{eq:evid1}
\end{align}

The two Gaussian distributions in the integral expression will generally be mismatched in dimensions  ($N_{\bphi}$ and $N_{\bphia}$). In order to facilitate the integration, we rewrite the $k$th kernel of the GMM in an expanded form as
\begin{equation} \label{eq:GaussianKernel}
\cN(\bphi|\bmuk,\bSigmak) = \cN\left( \begin{Bmatrix} \bphia \\ \bphima \end{Bmatrix} \Bigg|\begin{Bmatrix} \bmuak \\ \bmumak \end{Bmatrix}, \begin{bmatrix} \bSigmaak & \bCk  \\ \bCkT & \bSigmamak \end{bmatrix} \right) ,
\end{equation}
where $\bmuak$ and $\bSigmaak $ are the mean and covariance of the $k$th kernel pertaining to $\bphia$, likewise $\bmumak$ and $\bSigmamak$ are the mean and covariance of the $k$th kernel pertaining to $\bphima$, and $\bCk$ is the cross-covariance of $\bphia$ and $\bphima$. These entities are obtained from the construction of the GMM (whether by KDE, EM, or other means). Now, rewriting this joint distribution of $\bphia$ and $\bphima$ as the product of the conditional distribution of  $\bphima$ given $\bphia$ and the marginal distribution of $\bphia$, we obtain \cite{Sandhu2021}

\begin{subequations} \label{eq:GaussianKernel2}
\begin{align}
\cN(\bphi|\bmuk,\bSigmak) & = \cN(\bphima|\btilmumak,\btilSigmamak) \cN(\bphia|\bmuak,\bSigmaak), \label{eq:GaussianKernel2a} \\
 \btilmumak &= \bmumak + \bCkT (\bSigmaak)^{-1} (\bphia-\bmuak), \label{eq:GaussianKernel2b} \\
\btilSigmamak & = \bSigmamak - \bCkT (\bSigmaak)^{-1} \bCk  , \label{eq:GaussianKernel2c}
\end{align}
\end{subequations}
where $\btilmumak$ and $\btilSigmamak$ are the mean and covariance of the $k$th kernel, pertaining to $\bphima$ conditioned on a given value of $\bphia$. Consequently, the estimate of the model evidence becomes \cite{Sandhu2021}

 
 \begin{equation}  \label{eq:evidfinal}
\hat{\pr}(\bD|\balpha)=  \sum_{k=1}^{K} \ak \int \cN(\bphima|\btilmumak,\btilSigmamak) \cN(\bphia|\bmuak,\bSigmaak) \cN(\bphia|\bZO,\bA^{-1}) d\bphi,
\end{equation}
where the product of the two left-most Gaussians can be rewritten as

\begin{subequations}
\begin{align}
\cN(\bphia|\bmuak,\bSigmaak) \cN(\bphia|\bZO,\bA^{-1}) & =  \cN(\bmuak|\bZO,\bBak)  \cN(\bphia|\bmak \bPak) \\
\bmak &= \bmuak - \bSigmaak(\bBak)^{-1} \bmuak \label{eq:evidence_malpha} \\
\bPak & = \bSigmaak - \bSigmaak(\bBak)^{-1}\bSigmaak \label{eq:evidence_Palpha}\\
\bBak &= \bSigmaak + \bA^{-1}
\end{align}
\end{subequations}
where $\bmak$ and $\bPak$ are the mean and covariance of the $k$th kernel of the posterior of $\bphia$. Note that $\cN(\bmuak|\bZO,\bBak)$ is now independent of the parameters. Substituting and integrating over all parameters gives \cite{Sandhu2021}

\begin{align}
\hat{\pr}(\bD|\balpha)  & = \sum_{k=1}^{K} \ak \int \cN(\bphima|\btilmumak,\btilSigmamak) \cN(\bmuak|\bZO,\bBak)  \cN(\bphia|\bmak \bPak) d\bphi \nonumber \\
& =  \sum_{k=1}^{K} \ak  \cN(\bmuak|\bZO,\bBak) .
\end{align}

\subsection{Posterior parameter pdf}
\label{sec:parpostpdf}
As the estimate of the model evidence is now available in \eqref{eq:evidfinal}, it is now possible to obtain an estimate of the parameter posterior pdf $\hat{\pr}(\bphi|\bD,\balpha)$ making the same substitutions from above to obtain \cite{Sandhu2021}
\begin{equation} \label{eq:postpdf1}
 \hat{\pr}(\bphi|\bD,\balpha) = \sum_{k=1}^{K} \cfrac{\ak \cN(\bphi|\bmuk,\bSigmak) \cN(\bphia|\bZO,\bA^{-1}) }{\hat{\pr}(\bD|\balpha) }  .
\end{equation}

Substituting $\cN(\bphi|\bmuk,\bSigmak)$ from \eqref{eq:GaussianKernel2a} reduces \eqref{eq:postpdf1} to \cite{Sandhu2021}

\begin{align}
\hat{\pr}(\bphi|\bD,\balpha) &=\sum_{k=1}^{K} \cfrac{\ak \cN(\bphima|\btilmumak,\btilSigmamak) \cN(\bphia|\bmuak,\bSigmaak) \cN(\bphia|\bZO,\bA^{-1}) }{\hat{\pr}(\bD|\balpha) } \nonumber \\
& = \sum_{k=1}^{K}\underbrace{\left(\cfrac{ \ak \cN(\bmuak|\bZO,\bBak)  }{\sum_{r=1}^{K} a^{(r)} \cN(\bmu_{\alpha}^{(r)}|\bZO,\bB_{\alpha}^{(r)})}\right)}_{\wk}  \cN(\bphima|\btilmumak,\btilSigmamak)   \cN(\bphia|\bmak,\bPak) \label{eq:postpdf2} \\
 &= \sum_{k=1}^{K} \wk \cN\left( \begin{Bmatrix} \bphia \\ \bphima \end{Bmatrix} \Bigg|\begin{Bmatrix} \bmak \\ \bmmak \end{Bmatrix}, \begin{bmatrix} \bPak & \bDk  \\ \bDkT & \bPmak \end{bmatrix} \right) \label{eq:posterior_long}
\end{align}
where $0\leq\wk\leq1$, $\sum_k \wk =1$ is the weight coefficient of kernel $k$, and $\bmk$ and $\bPk$ are the mean and covariance of the $k$th kernel of the the posterior of $\bphi$, respectively. Note that $\bmak$ and $\bPak$ are known from  \eqref{eq:evidence_malpha}, and  \eqref{eq:evidence_Palpha}. Furthermore, $\bmmak$ and $\bPmak$ are the mean and covariance of the $k$th kernel of the posterior of $\bphima$ and $\bDk$ is the cross-covariance of $\bphia$ and $\bphima$. These entities are given by

\begin{subequations} 
 \begin{align}
\bmmak & = \bmumak - (\bCk)^T(\bBak)^{-1}\bmuak \\
\bPmak & = \bSigmamak - (\bCk)^{T}(\bBak)^{-1}\bCk \\
\bDk & = \bCk - \bSigmaak(\bBak)^{-1}\bmuak ((\bCk)^T(\bBak)^{-1}\bmuak)^T
\end{align}
\end{subequations}

\subsection{Gradient vector and Hessian matrix}
\label{sec:grad}
Let $J_{i}(\log \balpha)$ denote the $i$th element of the Jacobian vector $\bJ(\log\balpha)$. Differentiating the objective function in  \eqref{eq:cL} with respect to the  $\log(\alpha_i)$ gives \cite{Sandhu2021}

\begin{align}
J_i(\log\balpha) &=  \frac{\partial \cL(\log\balpha)}{\partial \log \alpha_i} =  \frac{\partial}{\partial \log \alpha_i}  \left\{  \log \hat{\pr}(\bD|\log\balpha) + \sum_{i=1}^{\Nalp} \left(r_i \log \alpha_i - s_i \alpha_i\right) \right\} \nonumber \\
& = \frac{\partial  \log \hat{\pr}(\bD|\log\balpha)}{\partial \log \alpha_i} + r_i - s_i \alpha_i \nonumber \; \\
& = \frac{1 }{\hat{\pr}(\bD|\log\balpha)} \frac{\partial }{\partial \log \alpha_i} \left\{ \sum_{k=1}^{K} \ak \cN(\bmuak|\bZO,\bBak)  \right\} + r_i - s_i \alpha_i  \nonumber \\
&=  \frac{1}{\hat{\pr}(\bD|\log\balpha)} \sum_{k=1}^{K} \ak \frac{\partial}{\partial \log\alpha_i} \left\{ \exp\left(\log\cN(\bmuak|\bZO,\bBak)\right)  \right\} + r_i - s_i \alpha_i \nonumber \\
&=  \sum_{k=1}^{K} \underbrace{\left(\frac{\ak \cN(\bmuak|\bZO,\bBak)}{\hat{\pr}(\bD|\log\balpha)}\right)}_{\wk} \underbrace{\left(\frac{\partial \log\cN(\bmuak|\bZO,\bBak)}{\partial \log \alpha_i}\right)}_{\vik} + r_i - s_i \alpha_i \\
& = \sum_{k=1}^{K} \wk \vik  + r_i - s_i \alpha_i, \label{eq:Jacobian}
\end{align}
where $\wk$ is known from \eqref{eq:postpdf2}, and factor $\vik$ is given as

\begin{align}
\vik &=  -\frac{1}{2}\left\{ -1 + \alpha_i \Piik  + \alpha_i  (\mik)^2 \right\} .
\end{align}

Now, let $H_{ij}(\log \balpha)$ denote element $(i,j)$ of the Hessian matrix $\bH(\log \balpha)$. Differentiating the $j$th element of the Jacobian in \eqref{eq:Jacobian} with respect to the $i$th element $\balpha$ gives \cite{Sandhu2021}

\begin{align}
H_{ij}(\log \balpha) &= \frac{\partial^2 \cL(\log \balpha)}{\partial \log\alpha_i \partial \log\alpha_j} = \frac{\partial  J_j(\log \balpha) }{\partial \log\alpha_i} = \frac{\partial}{\partial \log\alpha_i} \left\{  \sum_{k=1}^{K} \wk \vjk  + r_j - s_j \alpha_j \right\} \nonumber \\
& =  \sum_{k=1}^{K}\left\{ \wk \frac{\partial  \vjk }{\partial \log\alpha_i}  + \vjk \frac{\partial \wk }{\partial \log \alpha_i} \right\} -\delta_{ij} s_i\alpha_i \label{eq:Hessian},
\end{align}


with

\begin{align}
\frac{\partial  \vjk }{\partial \log\alpha_i} &=\alpha_i\alpha_j\left( \frac{(P_{ij}^{(k)})^2}{2} + m_i^{(k)} m_j^{(k)} P_{ij}^{(k)}  \right)  + \delta_{ij} \left( \vik -\frac{1}{2}  \right),\\
\frac{\partial  \wk }{\partial \log\alpha_i} &= \wk \left(\vik - \vibar \right),
\end{align}

where $\delta_{ij}$ is the Kronecker delta function, and $\vibar$ is the average of $\vik$ over all $K$ kernels.

\end{document}